\documentclass[final,5p,twocolumn,10pt]{elsarticle} 

\usepackage{lineno}
\usepackage[draft]{hyperref}
\usepackage{enumitem}
\modulolinenumbers[5]

\usepackage{amsmath}
\usepackage{amssymb}
\usepackage{stmaryrd}
\usepackage{amsthm}
\usepackage{amsfonts}
\usepackage{dsfont}
\usepackage{graphicx}
\usepackage{upgreek}
\usepackage{bm}
\usepackage{color}
\usepackage{float}
\usepackage{tikz-cd}
\usepackage{relsize}

\usepackage{mathtools}
\usepackage{algorithm,algpseudocode} 
\usepackage{todonotes}

\usepackage{subcaption}

\usepackage{tabu}
\usepackage{dblfloatfix} 

\biboptions{numbers,sort&compress}

\newcommand{\etal}{et al. }
\newcommand{\volSupp}{V_{S}}
\newcommand{\volSec}{V_\Gamma}
\newcommand{\volSuppm}{V_{S_m}}
\newcommand{\volSecm}{V_{\Gamma_m}}
\newcommand{\volSuppMax}{V_{{S}_{max}}}
\newcommand{\volSecMax}{V_{\Gamma_{max}}}
\newcommand{\buildDir}{\textbf{b}}
\newcommand{\buildDirSet}{B}

\newcommand{\cmm}{cm$^3$}

\newcommand{\bx}{{\mathbf{x}}}

\newcommand{\bk}{{\mathbf{k}}}

\newcommand{\bK}{{\mathbf{K}}}

\newcommand{\indic}{{\mathbf{1}}}

\newcommand{\vol}{\mathsf{vol}}

\newcommand{\R}{{\mathds{R}}}

\newcommand{\SO}[1]{{\mathrm{SO}(#1)}}

\theoremstyle{definition}

\newcommand{\eq}[1]{(\ref{#1})} 
\newcommand{\com}[1]{} 

\journal{Computers \& Graphics Special Issue on Computational Fabrication}

\begin{document}

\begin{frontmatter}

\title{Optimizing Build Orientation for Support Removal using Multi-Axis Machining}
   
\author{ Amir M. Mirzendehdel, Morad Behandish, and Saigopal Nelaturi}
\address{\rm Palo Alto Research Center (PARC), 3333 Coyote Hill Road, Palo Alto, California 94304  \vspace{-15pt}}

\begin{abstract}
 Parts fabricated by additive manufacturing (AM) are often fabricated first as a \emph{near-net} shape, a combination of intended nominal geometry and sacrificial support structures, which need to be removed in a subsequent post-processing stage using subtractive manufacturing (SM). In this paper, we present a framework for optimizing the build orientation with respect to \emph{removability} of support structures. In particular, given a general multi-axis machining setup and sampled build orientations, we define a Pareto-optimality criterion based on the total support volume and the ``secluded'' support volume defined as the support volume that is not \emph{accessible} by a given set of machining tools. Since total support volume mainly depends on the build orientation and the secluded volume is dictated by the machining setup, in many cases the two objectives are competing and their trade-off needs to be taken into account. 
 The accessibility analysis relies on the \emph{inaccessibility measure field} (IMF), which is a continuous field in the Euclidean space that quantifies the inaccessibility of each point given a collection of tools and fixturing devices. The value of IMF at each point indicates the minimum possible volumetric collision between objects in relative motion including the part, fixtures, and the tools, over all possible tool orientations and sharp points on the tool. 
 We also propose an automated support removal planning algorithm based on IMF, where a sequence of actions are provided in terms of the fixturing devices, cutting tools, and tool orientation at each step. In our approach, each step is chosen based on the maximal removable volume to iteratively remove accessible supports.
 The effectiveness of the proposed approach is demonstrated through benchmark examples in 2D and realistic examples in 3D.
\end{abstract}

\begin{keyword}	
	Build Direction Optimization \sep
	Support Structures\sep	
	Multi-Axis Machining \sep
	Accessibility Analysis \sep
	Spatial Planning \sep
	Hybrid Manufacturing
\end{keyword}

\end{frontmatter}

\section{Introduction} \label{sec_intro}

\begin{figure*} [h!]
		\centering
		\includegraphics[width=\linewidth]{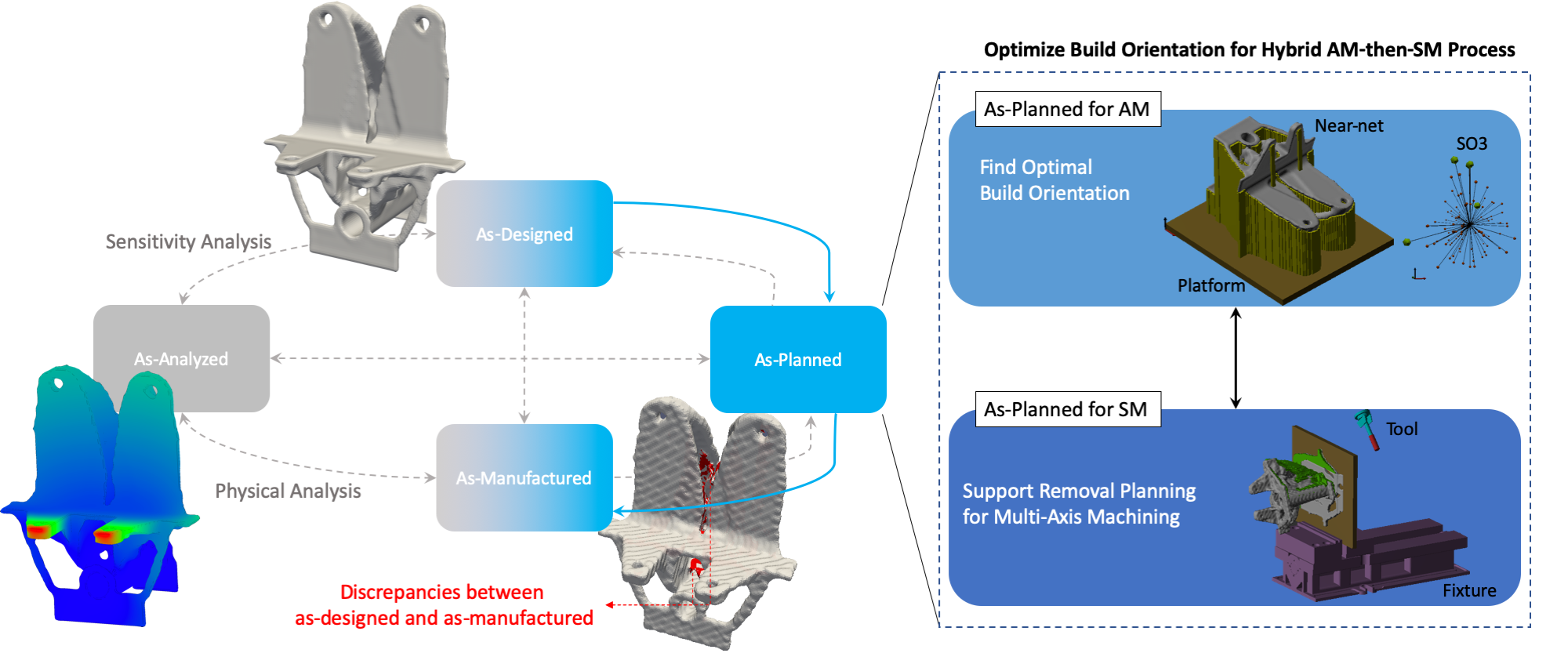}
		\caption{Different views of a product in design to manufacturing cycle. The four views of 1) as-designed, 2) as-planned, 3) as-manufactured, and 4) as-analyzed are interdependent and are traditionally reconciled through costly trial and error cycles. In this paper, we focus on the as-planned view for a hybrid AM-then-SM process such as metal-AM, where we find the optimal build orientation used in AM while considering the removability of support structures in a subsequent SM stage. Since we are not considering changing the as-designed model, there is no guarantee that \textit{all} supports will be removable and there may be secluded supports even at the optimal build orientation. The secluded regions will introduce discrepancies between the as-designed (intended) and as-manufactured (produced) parts.} \label{fig_views}
\end{figure*}

Additive manufacturing (AM) technologies are capable of fabricating geometrically complex parts by adding material layer-by-layer. The growing interest in AM, specifically metal-AM, stems from its ability to leverage geometric complexity to design high-performance light-weight designs for applications in aerospace, automotive, medical, etc. However, in most of metal-AM technologies such as powder-bed fusion, sacrificial support structures are needed in `overhanging' regions to dissipate excessive heat and ensure successful build of the near-net shape (part and support). In general, support structures directly add to the cost of AM parts by increasing the material usage, print time, and post-fabrication clean-up effort. Focusing on the latter, since the support structures need to be removed using subtractive machining (SM) such as milling, they need be `accessible' by the available machining tools in the presence of fixtures.
Thus, metal AM can be most effective in an AM-then-SM setting and a comprehensive manufacturing planning must consider both aspects. In other words, a cost-effective fabrication strategy should simultaneously minimize the costs associated with both printing the support and its clean-up. 

Figure \ref{fig_views} illustrates a design and manufacturing cycle for metal-AM parts where the four views of the part throughout the process --- namely, as-designed, as-planned, as-manufactured, and as-analyzed are interdependent. Lack of interoperability between representations, poor planning, and manufacturing limitations and uncertainties result in discrepancies between different views of the represented physical artifact, which are traditionally resolved through costly trial and error cycles. 

The focus of this paper is on improving the as-planned model w.r.t. support structures by considering the coupling between build orientation (AM) and multi-axis machining setup (SM). In particular, we are interested in finding the optimal build orientation such that 1) the volume of supports are minimized; and 2) the support structures are removable by a subsequent machining process.  Our method primarily relies on \textit{accessibility analysis} used in spatial planning. In particular, we extend the inaccessibility measure field (IMF) introduced in \cite{mirzendehdel2019exploring,mirzendehdel2020topology} to hybrid AM-then-SM processes. Previous work on topology optimization for multi-axis machining \cite{mirzendehdel2020topology} demonstrated the effectiveness of IMF as a continuous field in the Euclidean space that quantifies the collisions occurring between the part, fixture, and tool geometries to provide a penalty function to enforce accessibility by collision avoidance. We will briefly review the accessibility analysis and extend it to multiple fixtures in Section \ref{sec_imf}.   

There are numerous optimization strategies proposed to reduce the negative impact of support structures on the AM process, which typically focus on one or two of the views shown in Fig. \ref{fig_views}.
 Optimizing the build orientation is by far the most popular and there are numerous published articles on choosing the build orientation based on different criteria. Pandey \etal \cite{pandey2004optimal} optimized build time, surface quality, and support volume using a genetic algorithm. Jibin \cite{jibin2005determination} proposed a method for optimizing the build orientation based on multiple objectives, namely support volume, build time, and volumetric error between the as-designed and the as-manufactured parts. Nezhad \etal \cite{nezhad2010pareto} developed a Pareto-tracing algorithm to simultaneously minimize support volume and build time. Paul and Anand \cite{paul2015optimization} considered constraints on flatness and cylindricity errors while minimizing the support volume. Das \etal \cite{das2015optimum} presented a build direction optimization method to reduce tolerance errors and support volume. Umetani and Schmidt \cite{umetani2013cross} developed a framework to find optimal build orientation w.r.t. cross-sectional mechanical strength. Morgan \etal \cite{morgan2016part} presented a gradient descent algorithm for reducing support volume. Delfs \etal  \cite{delfs2016optimized} proposed a framework to predict surface quality of AM parts using a database of surface topographies and subsequently optimize the build orientation. Despite the large body of work on build orientation optimization, the feasibility and cost of post-processing for support removal is mainly ignored.

 Another way of reducing AM cost is generating efficient supports using tree-like structures \cite{vanek2014clever}, bridge scaffolding \cite{dumas2014bridging}, or topologically optimized supports \cite{allaire2020support,wang2020optimizing}. 
 
A third strategy to reduce the overall cost is to optimize the design such that it requires less support. This is achieved either by removing the supports entirely \cite{langelaar2016topology,guo2017self,qian2017undercut,zhang2018topology} or by considering the trade-off between multiple objective functions \cite{mirzendehdel2016support}. Considering the interlink between as-designed and as-planned views, Langelaar \cite{langelaar2018combined} and Wang and Qian \cite{wang2020simultaneous} proposed strategies to simultaneously optimize the part and build orientation w.r.t. support volume.  

Once the near-net shape is fabricated for an optimal build orientation, further post-process planning is required to specify a sequence of actions to remove the support volume either by manually detaching them or by machining them out using CNC milling. An automated support removal planning algorithm was proposed in \cite{nelaturi2019automatic}, where removability of supports was formulated based on accessibility of support beam ``dislocation features'' for columnar supports, defined by attachment points whose machining is sufficient to peel the columns off. The supports are recursively peeled off according to an optimal tool path obtained by solving a traveling salesman problem \cite{hoffman2013traveling}. In this paper, we propose an alternative approach using IMF and under the assumption that the entire support volume must be machined off, thereby enabling arbitrary support structures that are not necessarily columnar or attached at a finite number of points.. The details are explained in Section \ref{sec_planning}.       

As illustrated in Fig. \ref{fig_views}, our focus is on improving the as-planned view for hybrid metal AM-then-SM processes, where we find an optimized build orientation used in AM while also considering the removability of support structures in the SM stage. It is important to note that since we are not considering optimization of the as-designed model, there is inherently no guarantee that the entirety of the support volume will be removable and there may be secluded supports even at the optimized build orientation. Our method identifies such support regions and informs the designer so that the issues can be addressed. This is demonstrated in the results provided in Section \ref{sec_buildDir}. Future work will focus on incorporating support removability upfront in the design stage and coupling design optimization with build orientation optimization strategy developed in this paper.

\subsection{Contributions and Outline} \label {sec_outline}

In this paper, we present a framework for optimizing the build orientation w.r.t. removability of sacrificial support structures in hybrid metal AM-then-SM processes. In particular, given a general multi-axis machining setup and sampled build orientations we define a Pareto-optimality criterion based on the total support volume and the secluded support volume. Since total support volume mainly depends on the build orientation and the secluded volume is dictated by the machining setup (in each orientation) and the available tools, in many cases the two objectives are competing and their trade-off needs to be taken into account. 

The accessibility analysis relies on the concepts of inaccessibility measure field (IMF), which is a continuous field in the Euclidean space that quantifies the inaccessibility of each point given a collection of tools and fixturing devices. The value of IMF at each point indicates the minimum collision between stationary objects (part, platform, and fixtures) and cutting tools (tool holder and cutters), over all orientations and sharp points on the cutter. Further, the definition of IMF is based on well-established concepts widely used in spatial planning \cite{lozano1990spatial,nelaturi2015automatic}and enables realistic manufacturability analysis with geometrically complex tools and fixtures.

We also propose an automated support removal planning algorithm based on IMF, where a sequence of actions is provided in terms of the fixturing device, cutting tool, and orientation of the tool at each step. In our approach, each step is chosen based on 
the maximal volume that is removable by the combination of tools while avoiding collision with the machining setup to iteratively remove accessible support volume.

More specifically, the contributions of this paper are:

\begin{enumerate}
	\item extending the IMF definition to study accessibility of support structures in the near-net shape generated by an additive process;
	\item generalizing the IMF for a multi-axis machining setup with multiple fixturing devices/configurations; and
	\item developing a build orientation optimization for hybrid AM-then-SM processes based on Pareto-optimality of the total support volume and the secluded support volume.
\end{enumerate}

\section{Inaccessibility Measure Field} \label{sec_imf}
Here, we will briefly explain the accessibility analysis required for imposing the support accessibility constraint for multi-axis machining. The objective is to find regions of the support structure $S\subset \R^3$ that can be accessed by a cutting tool $T\subset \R^3$ without colliding with the workpiece $\Omega$, the platform $P\subset \R^3$, fixturing device $F\subset \R^3$.  Our approach is a generalization of the IMF proposed in \cite{mirzendehdel2020topology} for multiple fixturing devices. Let the tool assembly be $T = (H \cup K)$, where $H$ and $K$ are the tool holder and the cutter, respectively.

We assume that each tool assembly $T$ that can operate with six degrees of freedom (three translations and three rotations) \footnote{Note that if the tool is  axisymmetric, we need only two rotations to describe the tool orientation. Hence, we would only have five degrees of freedom.}.

We define the IMF, $f_\text{IMF}: \R^3 \to \R$, over the 3D design domain
 for each given tool assembly $T$ as the minimum collision volume between the stationary and moving objects over all sampled rotations in $\Theta_T \subset \SO{3}$ and sharp points $K \subseteq T$, where $\SO{3}$ is the group of different choices of sharp
points and available tool orientations $\Theta \subseteq \SO{3}$ \cite{mirzendehdel2020topology}:
\begin{equation}
f_\text{IMF}(\textbf{x};  O,N, T, K) := \min_{R \in \Theta_T} \min_{\textbf{k} \in K} ~\vol
\big[ O \cap (R, \textbf{x}) (T - \textbf{k}) \big], \label{eq_imf_1}
\end{equation}
where $R \in \Theta_T$ is a rotation matrix corresponding to an orientation, and $\textbf{x} \in \R^3$ is any query point inside, on the boundary, or outside the near-net shape  $N = \Omega \cup S$. $O = \Omega \cup (F \cup P)$ stands for the region in 3D space that is not subject to machining and must not be removed.  
There are two independent transformations in effect:
\begin{itemize}
	\item The shift $T \to (T - \bk)$ in \eq{eq_imf_1} is to try different ways to
	register the translation space with the design domain, by changing the local
	coordinate system to bring different sharp points to the origin.
	\item The rotation $(T - \bk) \to (RT - R\bk)$ followed by translation $(RT -
	R\bk) \to (RT - R\bk) + \bx$ bring the candidate sharp point (new origin) to
	the query point $\bx \in N$.
\end{itemize}
The same effect can be obtained by querying the convolution between the indicator functions of the stationary obstacle pointset $\indic_O$  and (rotated and reflected) moving pointset $\tilde{\indic}_{RT}$ at
$\textbf{t} := (\textbf{x }- R\bk)$ so that the rigid transformation $(R, \textbf{t})$ brings the
sharp point in contact with the query point: $(R,\textbf{ t})\bk = R\bk + \textbf{t} = R\bk +
(\textbf{x} - R\bk) = \textbf{x}$, as explained in \cite{mirzendehdel2020topology} in more detail. 
\begin{equation}
f_\text{IMF}(\bx; O,N, T, K) = \min_{R \in \Theta_T} \min_{\bk \in K}
~(\indic_{O} \ast \tilde{\indic}_{RT})(\bx - R\bk). \label{eq_imf_2}
\end{equation}
\noindent where $\ast$ is the convolution operator and  can be computed efficiently using fast Fourier transforms (FFTs). \\
We can extend \eq{eq_imf_2} to consider multiple machining tools. Given $n_T \geq 1$ available tool assemblies $T_i = (H_i \cup K_i)$ for $1 \leq
i \leq n_T$, we compute their combined IMF by applying another minimum operation over different tools to identify the tool(s) with the smallest volumetric interference at available orientations and sharp points:
\begin{equation}
f_\text{IMF}(\textbf{x};N, O) := \min_{1 \leq i \leq n_T} f_\text{IMF}(\textbf{x};N, O, T_i, K_i)
\label{eq_imf_3}
\end{equation}
in which $f_\text{IMF}(\textbf{x}; N,O, T_i, K_i)$ are computed from \eq{eq_imf_2}.
Finally, we can extend the IMF formulation of \eq{eq_imf_3} to multiple fixturing devices or configurations $F_j$ for $ 1 \le j \le n_F$, where each fixture results in a different obstacle $O_j = \Omega \cup (F_j \cup P)$. Thus,  the combined IMF for the entire machining setup can be computed as:
\begin{equation}
	f_\text{IMF}(\textbf{x};N) := \min_{1 \leq j \leq n_F} f_\text{IMF}(\textbf{x};N, O_j)
	\label{eq_imf_4}
\end{equation}
It is worthwhile noting that the $0-$level set of the IMF (i.e., the accessible region) is sufficient to distinguish accessible and inaccessible (i.e., secluded) support regions. However, the nonzero numerical values of the IMF over the secluded support regions are useful to quantify the degree of their inaccessibility. Such measures are useful to define continuous penalty functions for design optimization \cite{mirzendehdel2020topology}, which is beyond the scope of this paper.

Algorithm \ref{alg_imf} describes the procedure for computing IMF for multiple tools and fixtures using convolutions by FFT.

\begin{algorithm} [ht!]
	\caption{Compute $[{f}^{}_{\text{IMF}}]$.}
	\begin{algorithmic}
		\Procedure{IMF}{$[\indic_\Omega],[\indic_P], [\indic_{F_j}], [\indic_{H_i}],[\indic_{K_i}], \{\Theta_i\}$)}
		
		\State Initialize $[f_\text{IMF}] \gets [0]$
		\For{$j \gets 1$ to $n_F$}
		\State Define $[\indic_{O_j}] \gets [\indic_{\Omega}] + [\indic_{F_j}] + [\indic_P]$	\Comment{Union}
		\State Initialize $[f_\text{IMF}^{(j)}] \gets 0$ \Comment{IMF for all the tools}
		\For{$i \gets 1$ to $n_T$}
		\State Define $[\indic_{T_i}] \gets [\indic_{H_i}] + [\indic_{K_i}]$ 		\Comment{Implicit union}
		\State Define $V_{T_i} \gets \Call{Volume}{[\indic_{T_i}]}$ \Comment{Tool Vol.}

		\State Initialize $[\gamma_i] \gets [0]$ 
		\Comment{IMF for the $i^\text{th}$ tool}
		\ForAll{$R \in \{\Theta_i\}$}
		\State $[\indic_{RT_i}] \gets \Call{Rotate}{[\indic_{T_i}], R}$
		\State $[\tilde{\indic}_{RT_i}] \gets \Call{Reflect}{[\indic_{RT_i}]}$
		\State $[g] \gets \Call{Convolve}{[\indic_{O_j}], [\tilde{\indic}_{R T_i}]}$
		\State $[g] \gets [g]/V_{T_i}$ \Comment{Normalize}
		\ForAll{$\bk \in \Call{Support}{[\indic_{K_i}]}$}
		\State $[h] \gets \Call{Translate}{[g], -R \bk}$
		\State $[\gamma_i] \gets \min([\gamma_i], [h])$
		\Comment{ Over sharp pts}
		\EndFor
		\EndFor
		\State $[f_\text{IMF}^{(j)}] \gets \min([f_\text{IMF}^{(j)}], [\gamma_i])$
		\Comment{Over all tools}
		\EndFor
		\State $[f_\text{IMF}] \gets \min([f_\text{IMF}], [f_\text{IMF}^{(j)}])$
		\Comment{Over all fixtures}
		\EndFor
		\State\Return{($[f_\text{IMF}]$)}
		\EndProcedure 
	\end{algorithmic} \label{alg_imf}
\end{algorithm}

\begin{figure}
	\centering
	\includegraphics[width=0.95\linewidth]{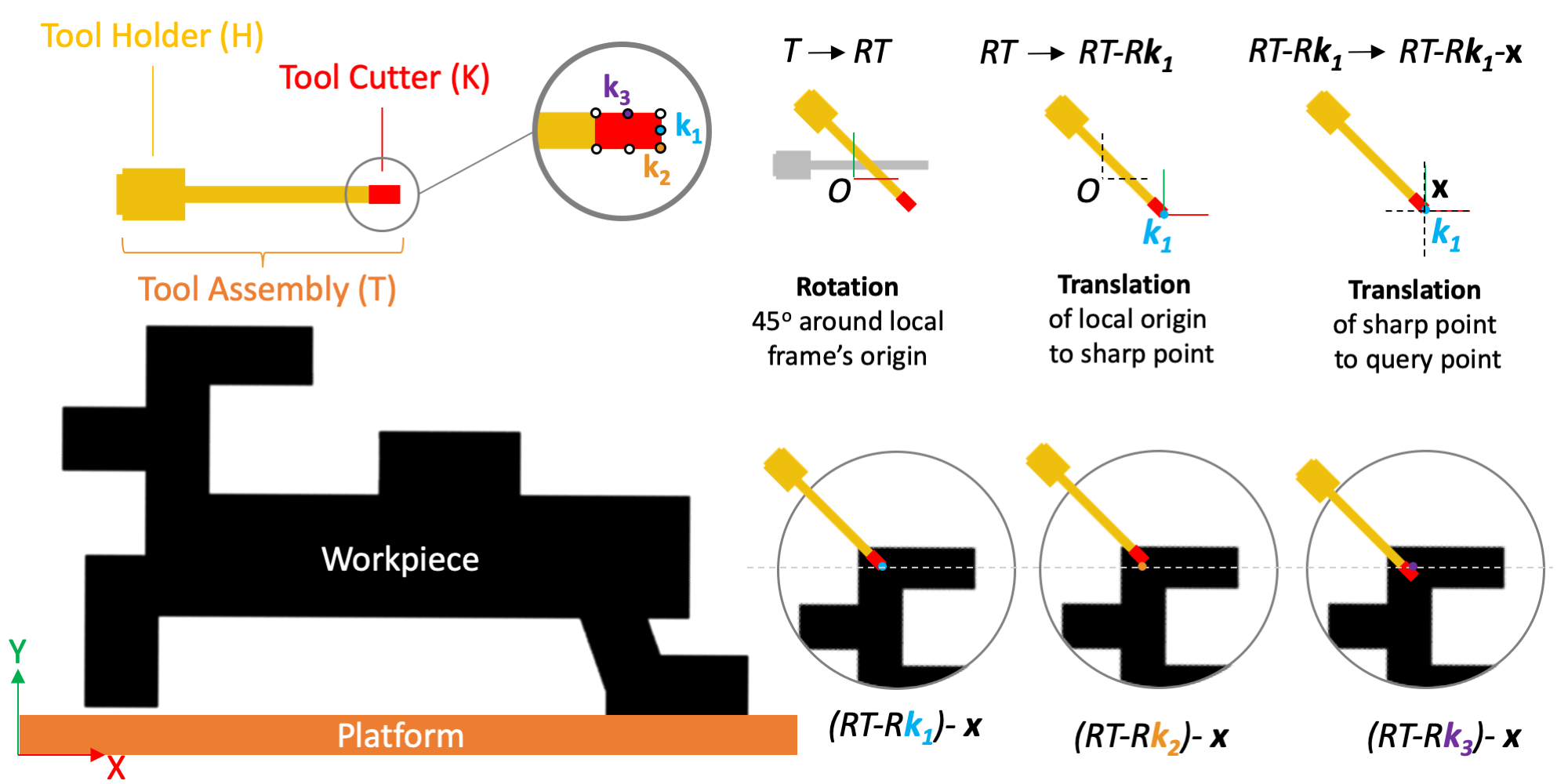}
	\caption{Workpiece is fabricated by AM on a platform and is post-processed for support removal by multi-axis machining given the cutting tool holder, cutter, and orientations.}
	\label{fig_design}
\end{figure}

\begin{figure} [ht!]
	\begin{subfigure}[t]{0.48\linewidth}
		\centering
		\includegraphics[width=0.85\linewidth]{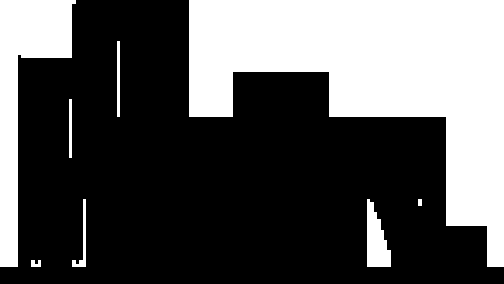}
		\caption{Near-net shape at build orientation $\buildDir = (0,+1)$.}
	\end{subfigure}%
	\begin{subfigure}[t]{0.5\linewidth}
		\centering
		\includegraphics[width=0.99\linewidth]{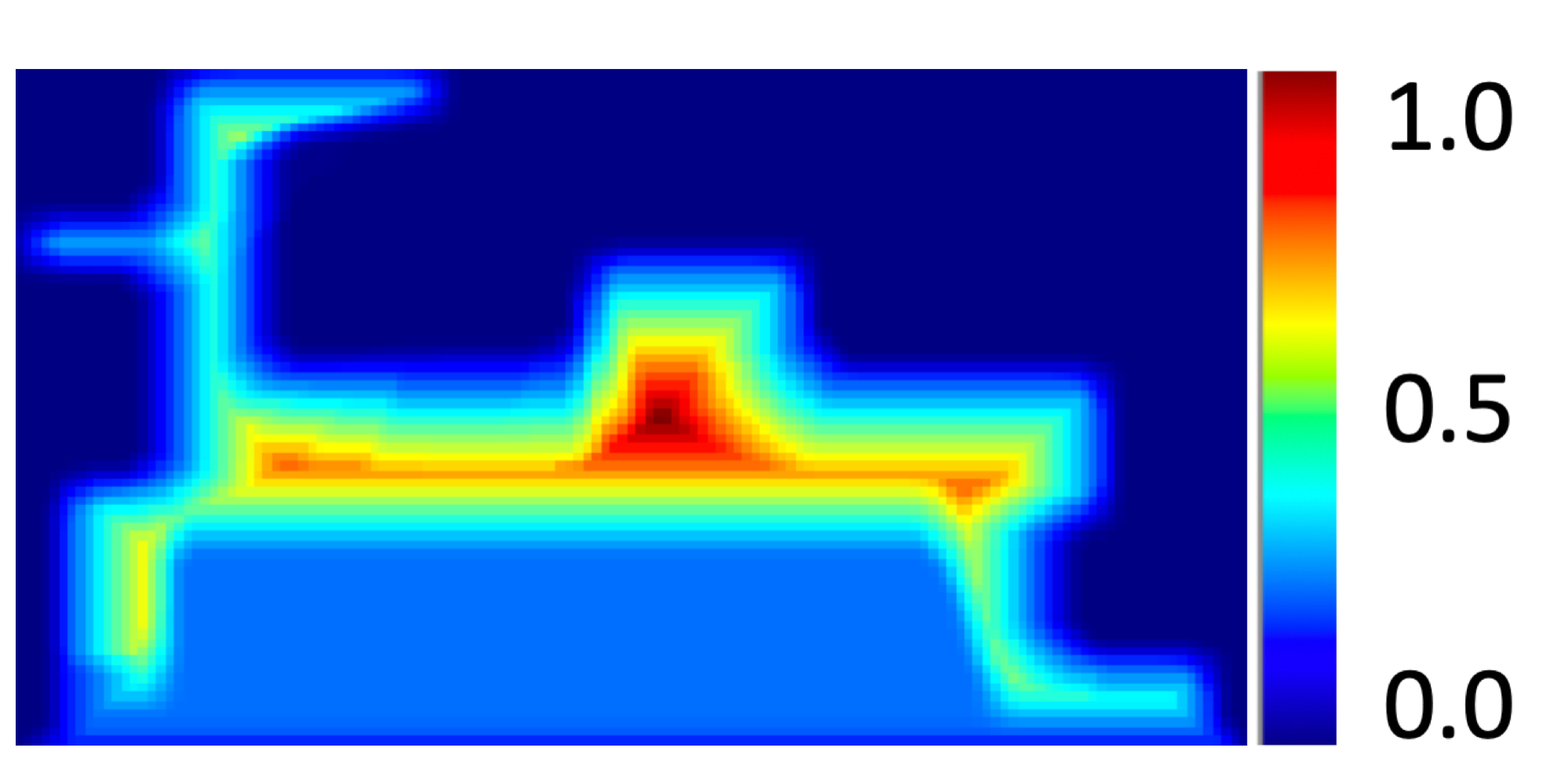}
		\caption{IMF over the near-net shape.}
	\end{subfigure}%
	
	\begin{subfigure}[t]{0.5\linewidth}
		\centering
		\includegraphics[width=0.99\linewidth]{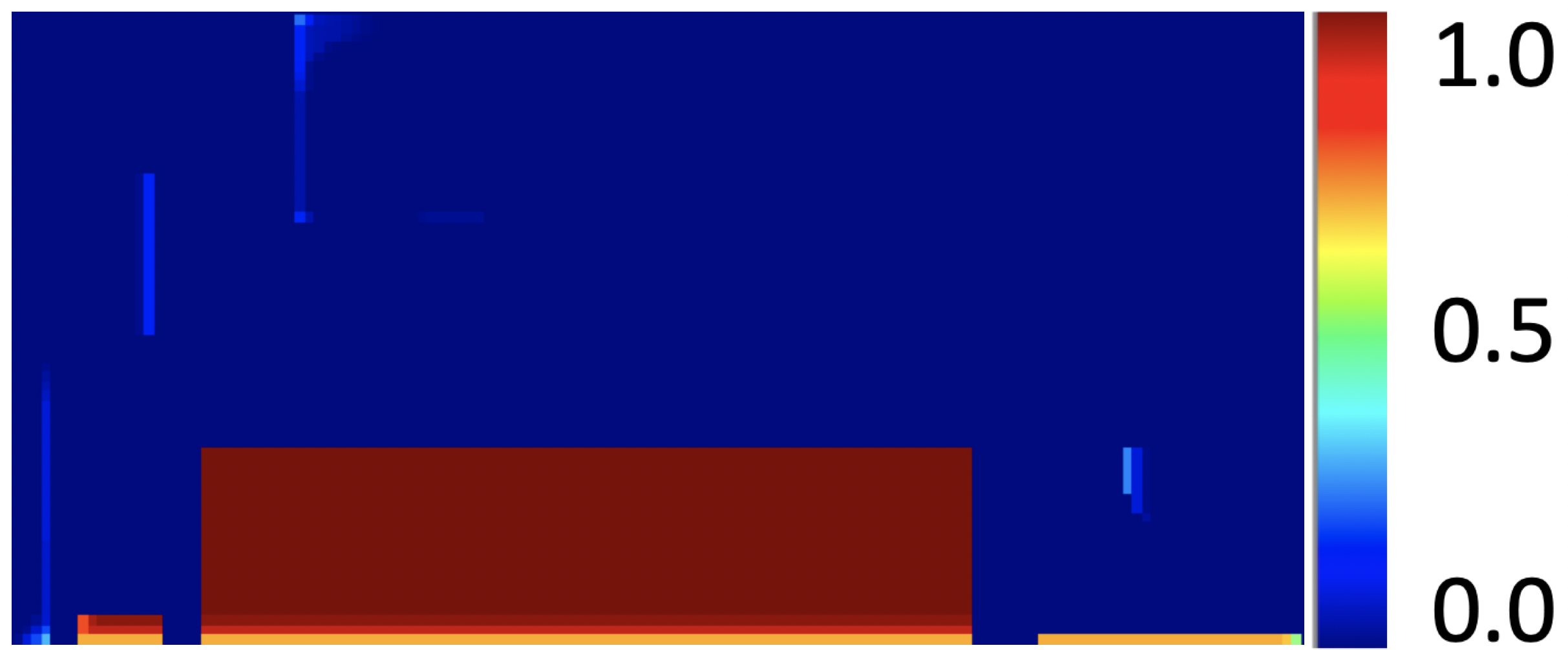}
		\caption{IMF over supports.}
	\end{subfigure}%
	\begin{subfigure}[t]{0.48\linewidth}
		\centering
		\includegraphics[width=0.8\linewidth]{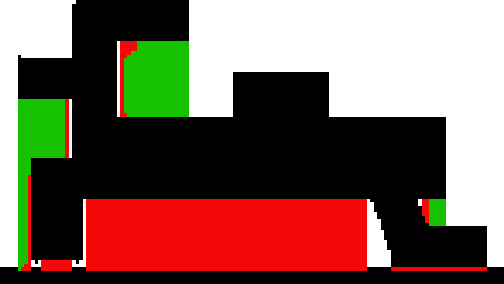}
		\caption{Accessible (green) and Inaccessible (red) supports.}
	\end{subfigure}%
	\caption{Accessibility analysis for supports at $\buildDir = (0,+1)$. Obstacle is the union of the design and the platform. } \label{fig_accAnalysis}
\end{figure}

\begin{figure} [h!]
	\begin{subfigure}[t]{0.5\linewidth}
		\centering
		\includegraphics[width=0.9\linewidth]{fig/designWsupp_0.png}
		\caption{Build orientation with $\theta = 0^\circ$.}
	\end{subfigure}%
	\begin{subfigure}[t]{0.5\linewidth}
		\centering
		\includegraphics[width=0.9\linewidth]{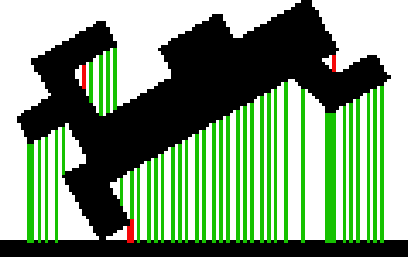}
		\caption{Build orientation with $\theta = 30^\circ$.}
	\end{subfigure}%
	
	\begin{subfigure}[t]{0.5\linewidth}
		\centering
		\includegraphics[width=0.7\linewidth]{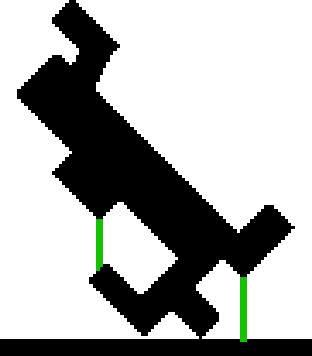}
		\caption{Build orientation with $\theta = 135^\circ$.}
	\end{subfigure}%
	\begin{subfigure}[t]{0.5\linewidth}
		\centering
		\includegraphics[width=0.9\linewidth]{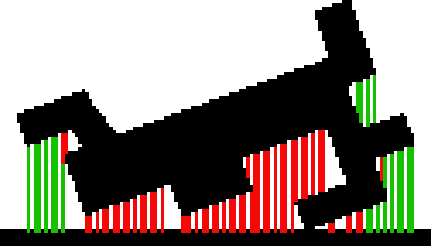}
		\caption{Build orientation with $\theta = 200^\circ$.}
	\end{subfigure}%
	\caption{Accessible and inaccessible support structures at different build angles.} \label{fig_buildAngles}
\end{figure}

The accessibility analysis for the example of Fig. \ref{fig_design} with build direction along Y axis, $\buildDir := (0,+1)$, is illustrated in Fig. \ref{fig_accAnalysis}. \\
A general build direction in 2D is denoted by $\buildDir= R(\theta) (0,+1)$, where $R(\theta)$ is the rotation matrix in 2D by an angle $\theta$. Figure \ref{fig_buildAngles} shows accessible and inaccessible supports at four different build directions for overhang angle threshold $\alpha = 45^\circ$.

\section{Build Orientation Optimization} \label{sec_buildDir}

In this section, we describe the build orientation optimization method w.r.t. 1) support volume and 2) removability of supports under the assumption that the near net shape is fixturable in every orientation. In the proposed method, we find the Pareto-optimal build orientation $\buildDir^*_{}$ considering two objectives, where volume of the entire support $\volSupp(\Omega)$ and the secluded supports $\volSec(\Omega)$ are to be minimized. 
 
 By definition, a Pareto-optimal solution gives the best trade-off between multiple quantities of interest. In other words, further improvement of one objective worsens at least one other. A set of Pareto-optimal solutions is called a Pareto frontier, which can be quite useful in engineering  applications  where  multiple  optimized  designs can be explored so that the one with the most suitable trade-off is chosen. 
 
 $\volSupp(\Omega)$ depends on the build orientation while $\volSec(\Omega)$ depends on the machining setup and available fixturing configurations, tools and orientations. Therefore, in many cases an orientation with less support volume does not necessarily result in a smaller secluded support region. Thus, we need to define a parameter that captures these objectives according to some user-specified level of importance for each. To this end, we define $\xi$ at the $m^{th}$ build orientation $\buildDir_m$ as weighted sum of the support volume $\volSupp(\Omega,\buildDir_m)$ and secluded support volume $\volSec(\Omega,\buildDir_m)$:
 
\begin{equation}
\xi =(1-w_{acc})\dfrac{\volSupp(\Omega,\buildDir_m)}{\volSuppMax} + w_{acc} \dfrac{\volSec(\Omega,\buildDir_m)}{\volSecMax}
\label{eq_weigthedVal}
\end{equation}

where $0\le w_{acc}\le 1$ is the user-specified weight for considering accessibility. $\volSupp(\Omega)$ and $\volSec(\Omega)$ in \eq{eq_weigthedVal} are normalized w.r.t. $\volSuppMax$ and $\volSecMax$ denoting the maximum values of support volume and secluded volume over all given build orientations, respectively. In this paper, we sample the 3D rotation group $\SO{3}$ to obtain the set of candidate build orientations. However, there are many gradient-free or stochastic optimization methods \cite{davis1987genetic} such as simulated annealing \cite{mahfoud1995parallel} that can be used to sample $\SO{3}$ more efficiently \cite{nelaturi2013solving} or converge to global optimum.

Algorithm \ref{alg_buildDirOpt} describes the build direction optimization w.r.t. support volume and support accessibility.

\begin{algorithm} [ht!]
	\caption{Optimize Build Orientations $\buildDirSet^*$.}
	\begin{algorithmic}
		\Procedure{Optimize}{$w_{acc},\Omega,P, [\indic_F],[\indic_H],[\indic_K],n_b,n_{b^*}, \lambda$}
		\State $\buildDirSet \gets \Call{Sample}{n_b}$ \Comment{Sampling of $\SO{3}$}
		\State Initialize $\{\volSuppm\}, \{\volSecm\},\{\xi_m\}$ \Comment{Vectors of  size $m$}
		\For{$m \gets 1$ to $n_b$}
		\State $\Omega_m\gets \Call{Rotate}{\Omega, b_m}$ 
		\State $S \gets \Call{SuppGen}{\Omega_m}$ \Comment{Generate support}
		\State $[{f}_\text{IMF}] \gets\Call{IMF}{\indic_{\Omega_m},\cdots}$ \Comment{Compute IMF}
		\State $\indic_\Gamma \gets [\indic_S] \cdot [{f}_\text{IMF} > \lambda] $ \Comment{Secluded supports}
		\State $\volSuppm \gets \Call{Volume}{S}$ \Comment{Compute volume}
		\State $\volSecm \gets \Call{Volume}{\Gamma}$
		\EndFor
		\State $\volSuppMax \gets \Call{Max}{[\volSuppm]}$ \Comment{Find maximum value}
		\State $\volSecMax \gets \Call{Max}{[\volSecm]}$
		\For{$m \gets 1$ to $n_b$}
		\State $\xi_m \gets w_{acc}\dfrac{\volSuppm}{\volSuppMax} + (1-w_{acc}) \dfrac{\volSecm}{\volSecMax}$
		\EndFor
		\State $\buildDirSet^*_{} \gets \Call{Sort}{[\xi_m],n_{b^*}}$ \Comment{ Get top $n_{b^*}$ solutions}
		\State\Return{($\buildDirSet^*_{}$)}
		\EndProcedure 
	\end{algorithmic} \label{alg_buildDirOpt}
\end{algorithm}

\subsection{Benchmark in 2D: Impact of Overhang Angle}
\begin{figure*}[!ht]
	\begin{subfigure}[t]{0.8\linewidth}
		\centering
		\includegraphics[width=0.8\linewidth]{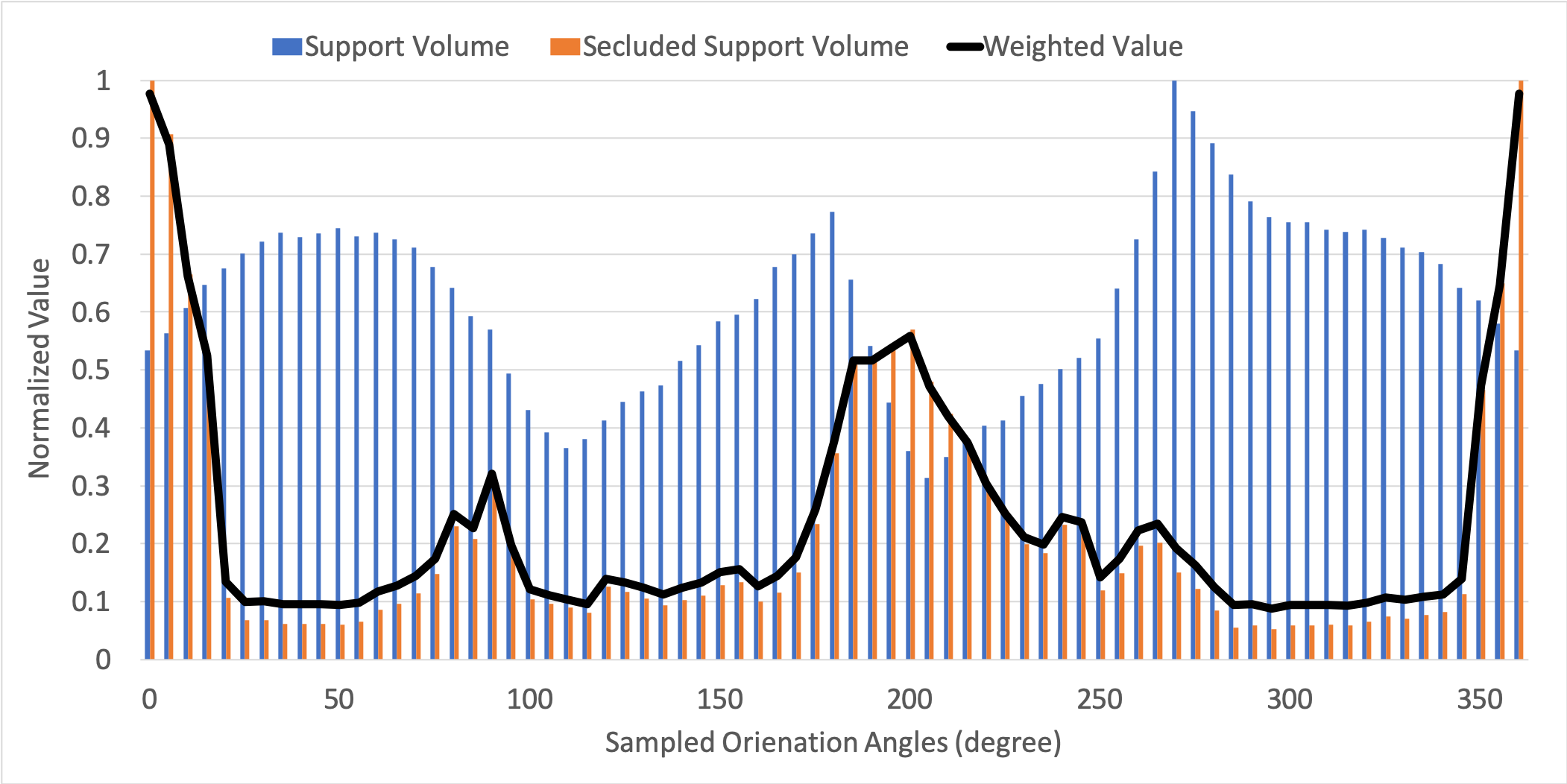}
		\caption{Sampled build orientations with corresponding support volume, secluded volume, and weighted Pareto-optimality value. }
		\label{fig_volFracs90}
	\end{subfigure}
	\begin{subfigure}[t]{0.18\linewidth}
		\centering
		\includegraphics[width=0.8\linewidth]{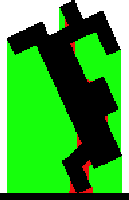}
		\caption{An optimized build orientation at a conservative $90^\circ$ overhang angle.}
		\label{fig_design295_90}
	\end{subfigure}
	\caption{Optimizing build orientation with overhang angle $\alpha = 90^\circ$ and Pareto accessibility weight $w_{acc} = 0.95$.  There are no sampled build orientations with fully removable support structures and the Pareto-optimal build orientation is $\buildDir^*_{} = 295^\circ$. }
\end{figure*} 

\begin{figure*}[!ht]
	\begin{subfigure}[t]{0.8\linewidth}
		\centering
		\includegraphics[width=0.80\linewidth]{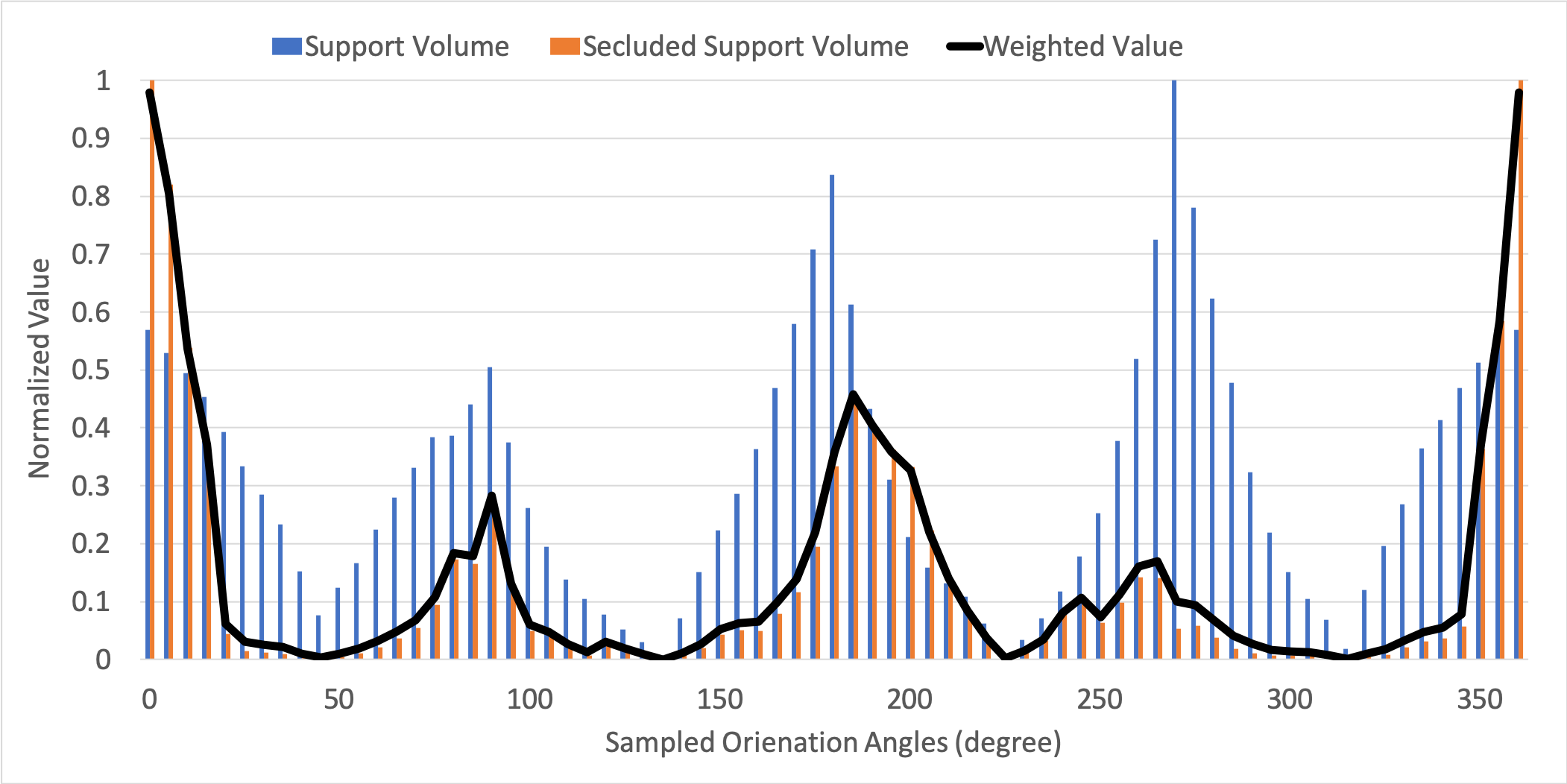}
		\caption{Sampled build orientations with corresponding support volume, secluded volume, and weighted Pareto-optimality value.  }
		\label{fig_volFracs45}
	\end{subfigure}
	\begin{subfigure}[t]{0.18\linewidth}
		\centering
		\includegraphics[width=0.99\linewidth]{fig/designWsupp_135}
		\caption{An optimized build orientation at $45^\circ$ overhang angle.}
		\label{fig_design135_45}
	\end{subfigure}
	\caption{Optimizing build orientation with overhang angle $\alpha = 45^\circ$ and Pareto accessibility weight $w_{acc} = 0.95$. There are 3 build orientations with fully removable structures, the Pareto-optimal solution is at the build angle $135^\circ$.}
\end{figure*} 
Overhang angle $\alpha$ is an important factor in optimizing the build orientation and can have a significant impact on both support volume and location of supports. In this section, we consider the part of Fig. \ref{fig_design} as a benchmark example in 2D, where optimized build orientations for two typical overhang angles $\alpha = 90^\circ$ and $\alpha = 45^\circ$ are considered. The set of build orientations $\buildDirSet\subset \R^{2}$ with 72 sampled orientations $\theta \in \SO{2}$ are:
\begin{equation}
\buildDirSet= \{ R(\theta) (0,+1) ~|~ \theta = 2\pi i/72 ~\text{for}~ i = 0, 1, \ldots, 71\}
\end{equation}
where $R(\theta)$ represents a planar rotation matrix corresponding to rotation angle of $\theta$.

The multi-axis machining tool used for support removal is shown in Fig. \ref{fig_design}. The set of tool directions with 36 sampled orientations is:
\begin{equation}
	\Theta= \{ R(\theta) ~|~ \theta = 2\pi i/36 ~\text{for}~ i = 0, 1, \ldots, 35\}
\end{equation}

The accessibility analysis, support generation, and optimization are performed at $256\times256$ grid resolution. For this example, we choose $w_{acc} = 0.95$ to find the best build orientation w.r.t. support removability.

Figure \ref{fig_volFracs90} shows the volume of support and inaccessible supports at different build orientations for the conservative overhang angle of $\alpha = 90^\circ$ and $\lambda = 0.01$. There are no orientations with fully accessible/removable support structures. The orientation with minimum inaccessible support volume is $\buildDir^*_{} =  R(295^\circ)(0,+1)$ as shown in Fig. \ref{fig_design295_90} with ${\volSupp(\Omega,\buildDir_m)}/{\volSuppMax}= 0.76$ and ${\volSec(\Omega,\buildDir_m)}/{\volSecMax}= 0.06$. The Pareto-optimality value $\xi$ minimized is $0.088$.

To highlight the impact of overhang angle $\alpha$ on the optimized build orientation, Fig. \ref{fig_volFracs45} shows the volume of support and inaccessible supports at different build orientations. There are no inaccessible support volumes at build angles $45^\circ, 135^\circ, 315^\circ$. Among them, the Pareto-optimal solution is $\buildDir^*_{} =  R(135^\circ)(0,+1)$ as shown in Fig. \ref{fig_design135_45} with ${\volSupp(\Omega,\buildDir_m)}/{\volSuppMax}= 0.007$ and ${\volSec(\Omega,\buildDir_m)}/{\volSecMax}= 0.0$. The Pareto-optimality value is very close to zero, $\xi = 0.0004$. 

\subsection{Support Bracket: Impact of Accessibility Weight}
\begin{figure} [h!]
	\centering
	\includegraphics[width=0.99\linewidth]{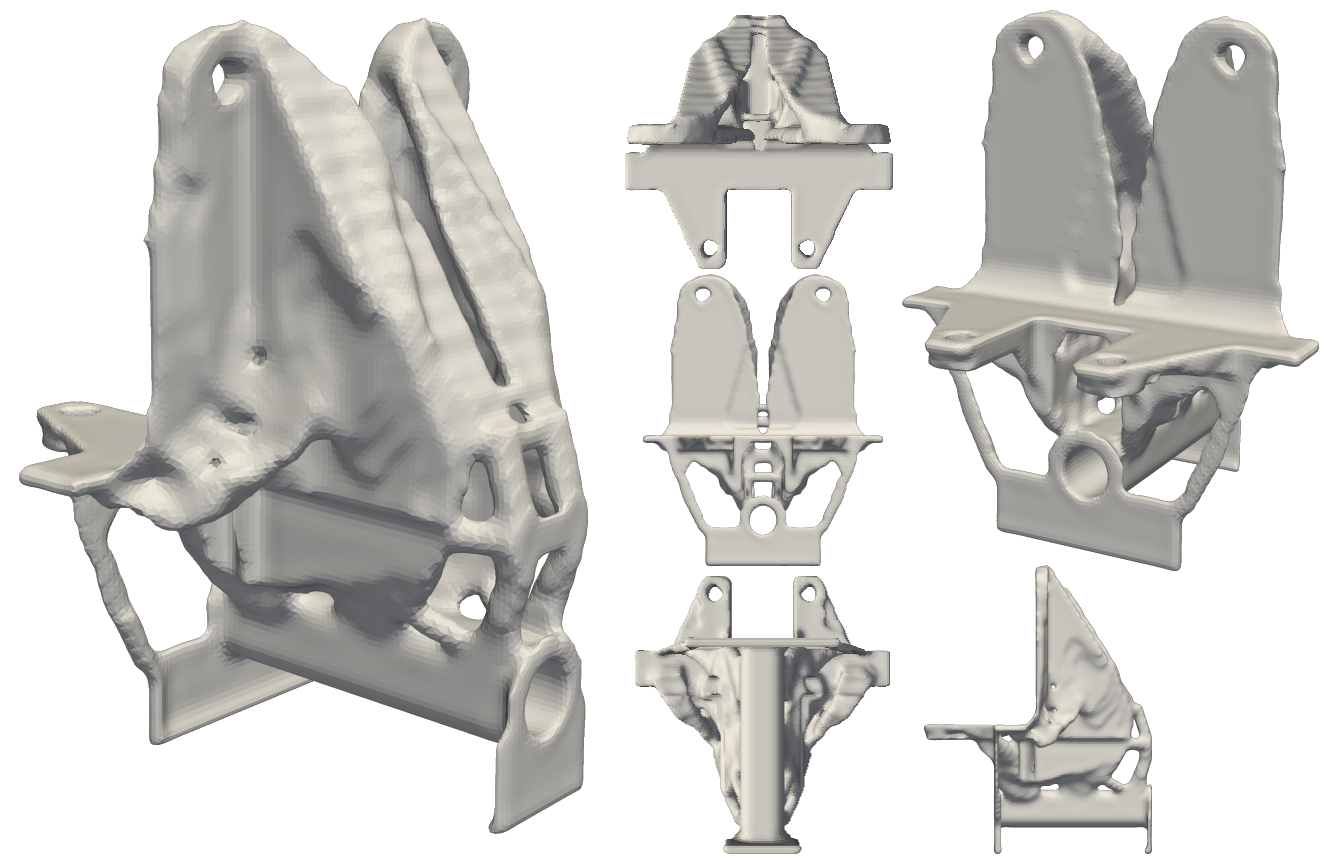}
	\caption{Topologically optimized support bracket \cite{mirzendehdel2020topology}.} \label{fig_SuppBracketGeom}
\end{figure}

\begin{figure} [ht!]
	\begin{subfigure}[t]{0.5\linewidth}
		\centering
		\includegraphics[width=0.9\linewidth]{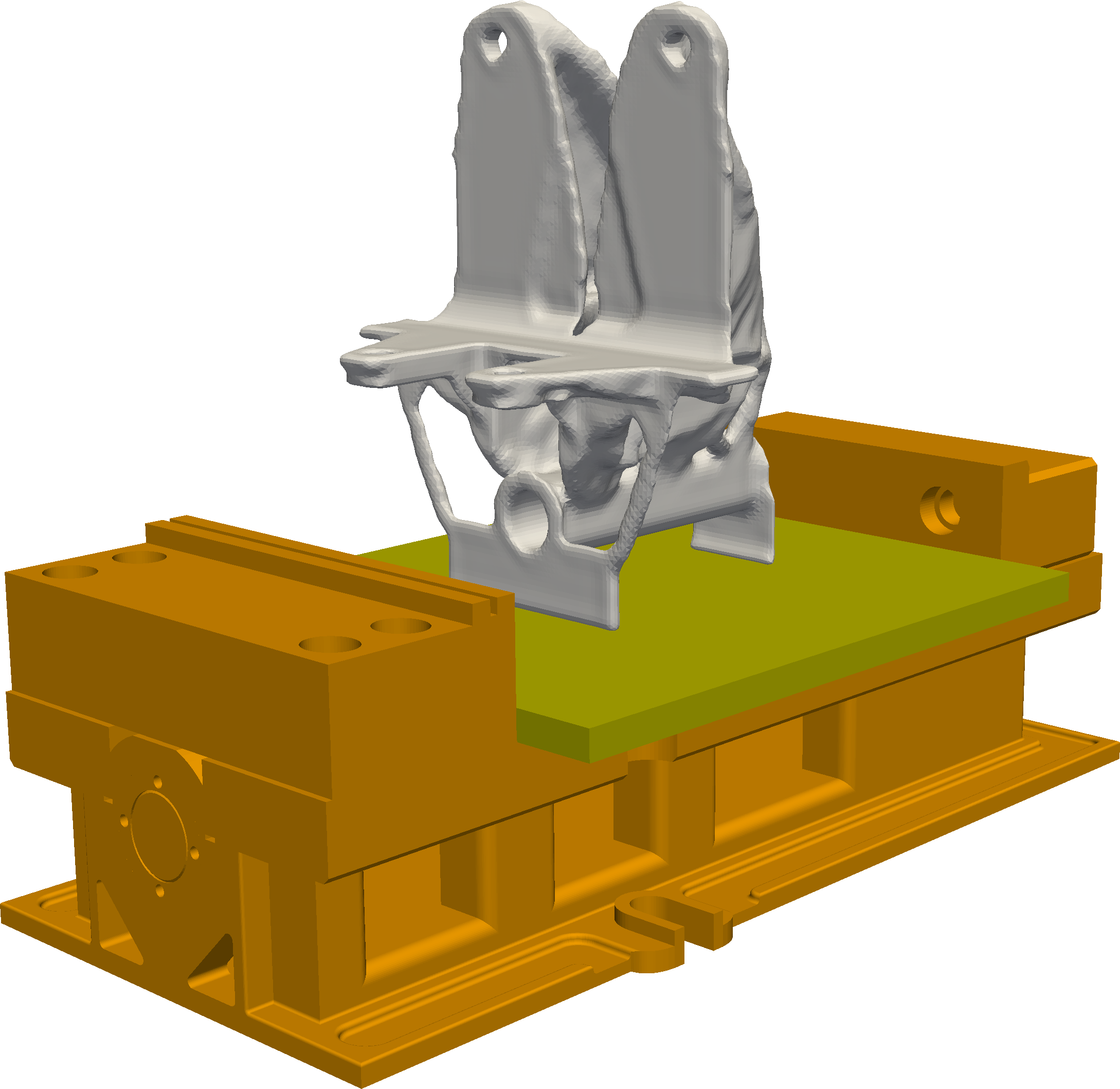}
		\caption{Fixture 1.}
	\end{subfigure}%
	\begin{subfigure}[t]{0.5\linewidth}
		\centering
		\includegraphics[width=0.9\linewidth]{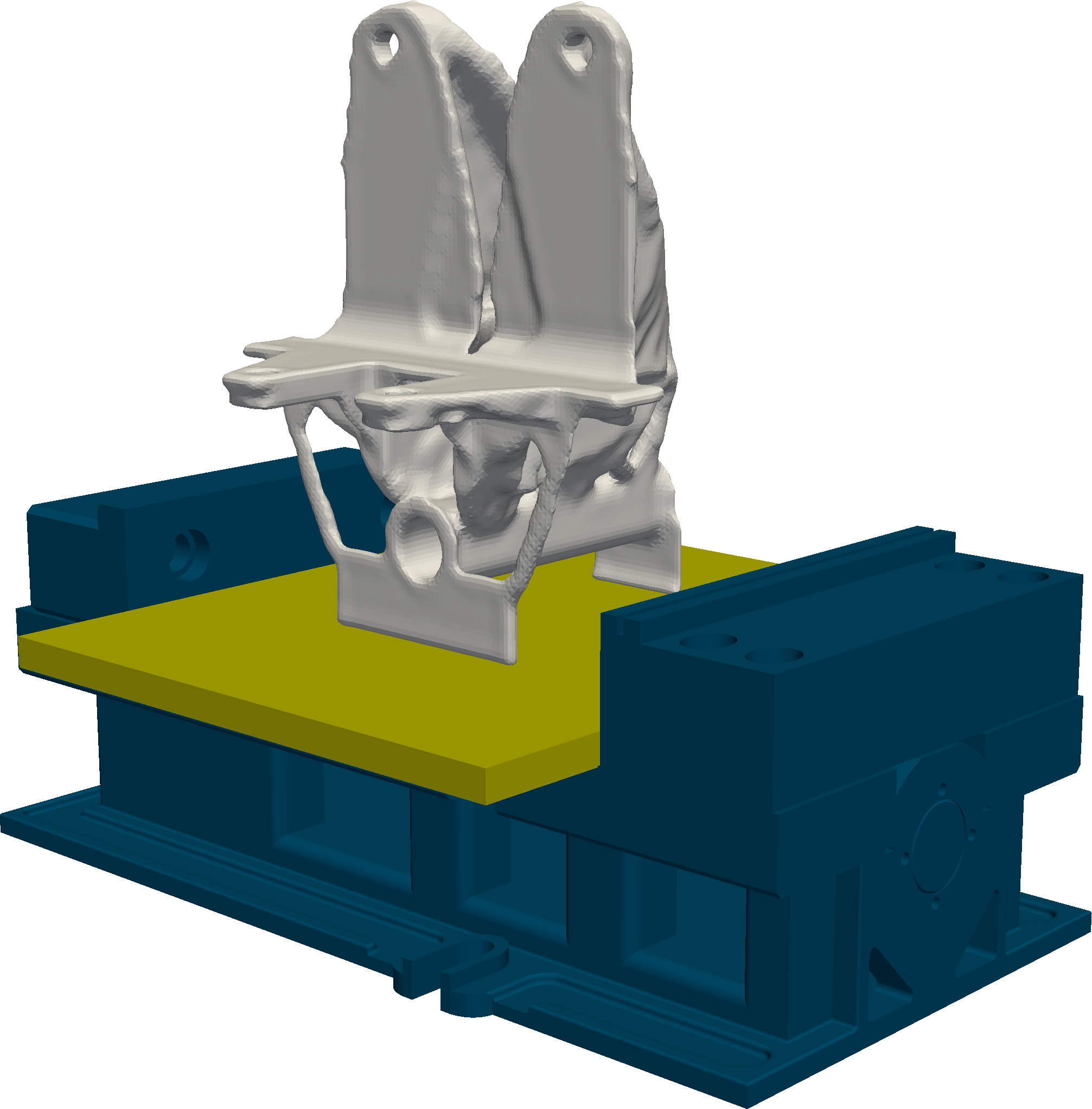}
		\caption{Fixture 2.}
	\end{subfigure}%
	
	\begin{subfigure}[t]{0.5\linewidth}
		\centering
		\includegraphics[width=0.9\linewidth]{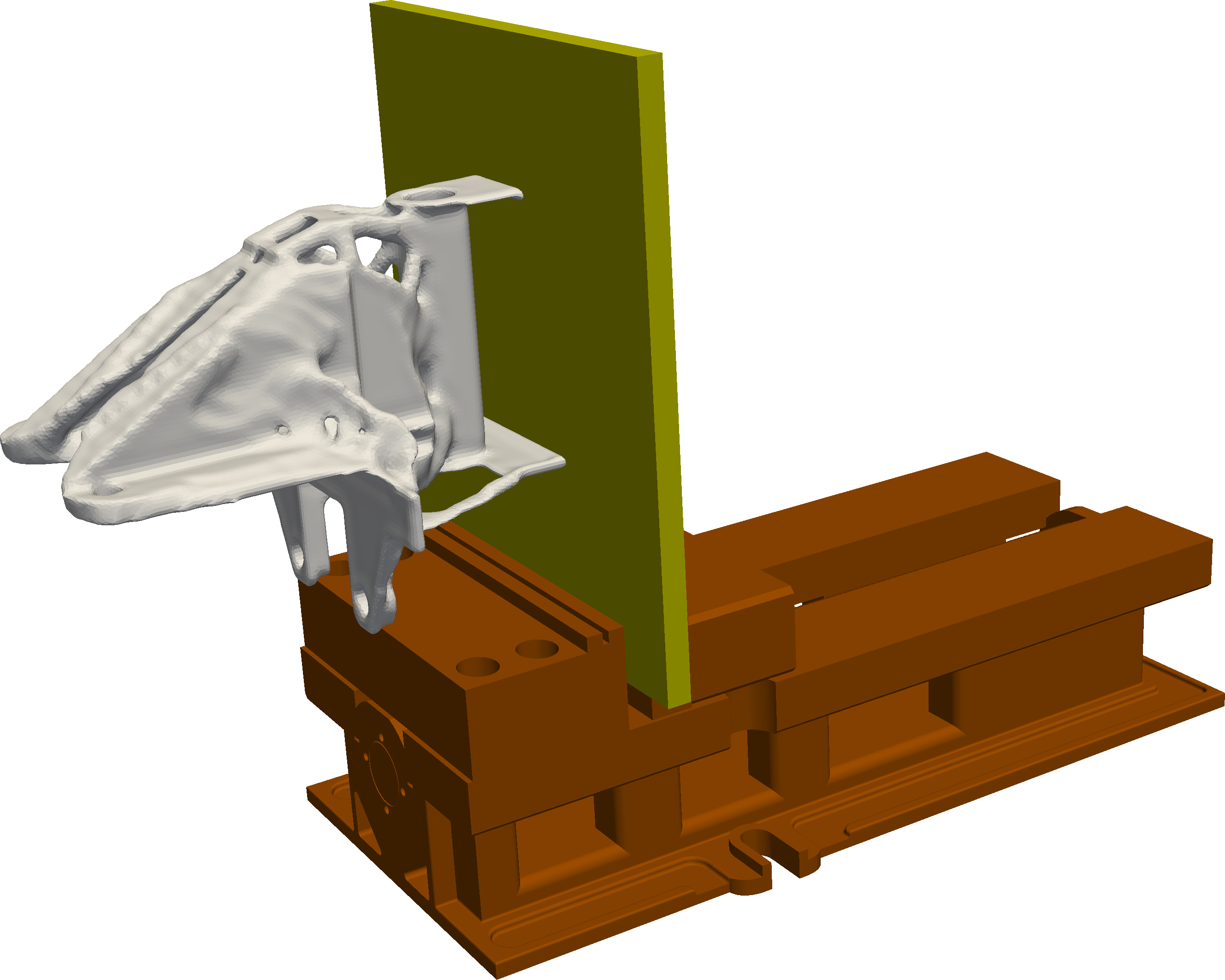}
		\caption{Fixture 3.}
	\end{subfigure}%
	\begin{subfigure}[t]{0.5\linewidth}
		\centering
		\includegraphics[width=0.9\linewidth]{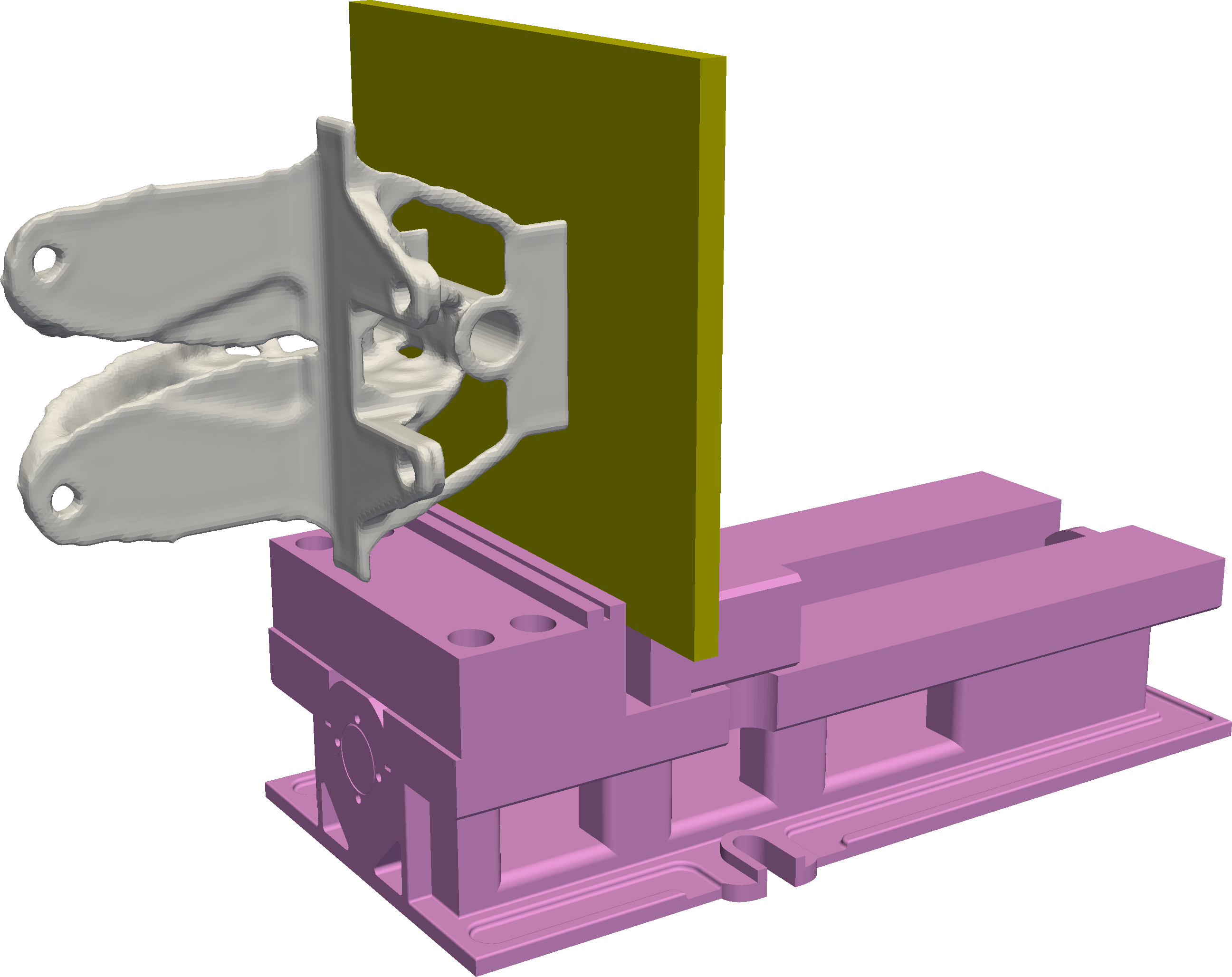}
		\caption{Fixture 4.}
	\end{subfigure}%
	\caption{Four fixturing configurations.} \label{fig_fixtures}
\end{figure}

\begin{figure} [ht!]
	\begin{subfigure}[t]{\linewidth}
		\centering
		\includegraphics[width=0.5\linewidth]{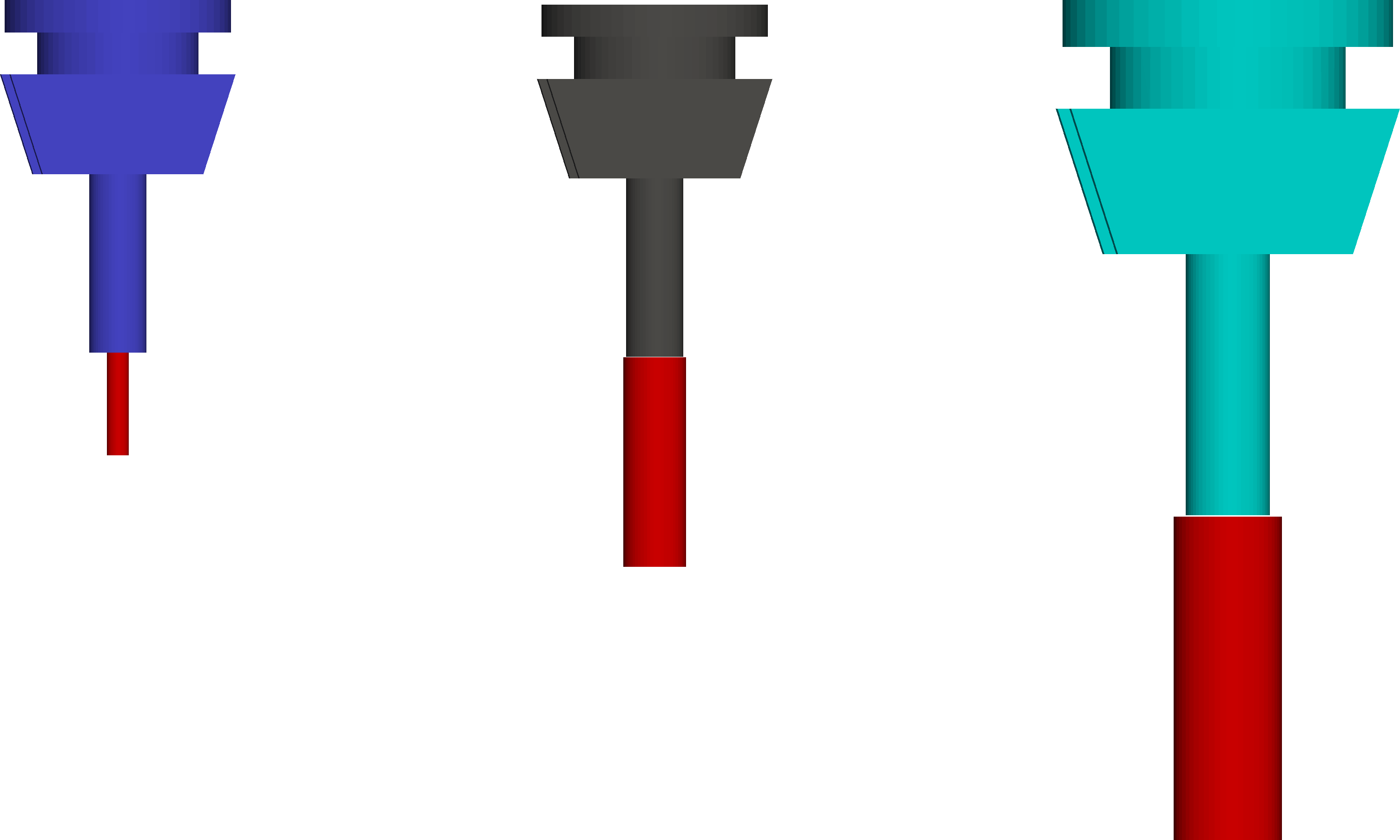}
		\caption{3 different machining tools.}
	\end{subfigure}%
	
	\begin{subfigure}[t]{0.33\linewidth}
		\centering
		\includegraphics[width=0.9\linewidth]{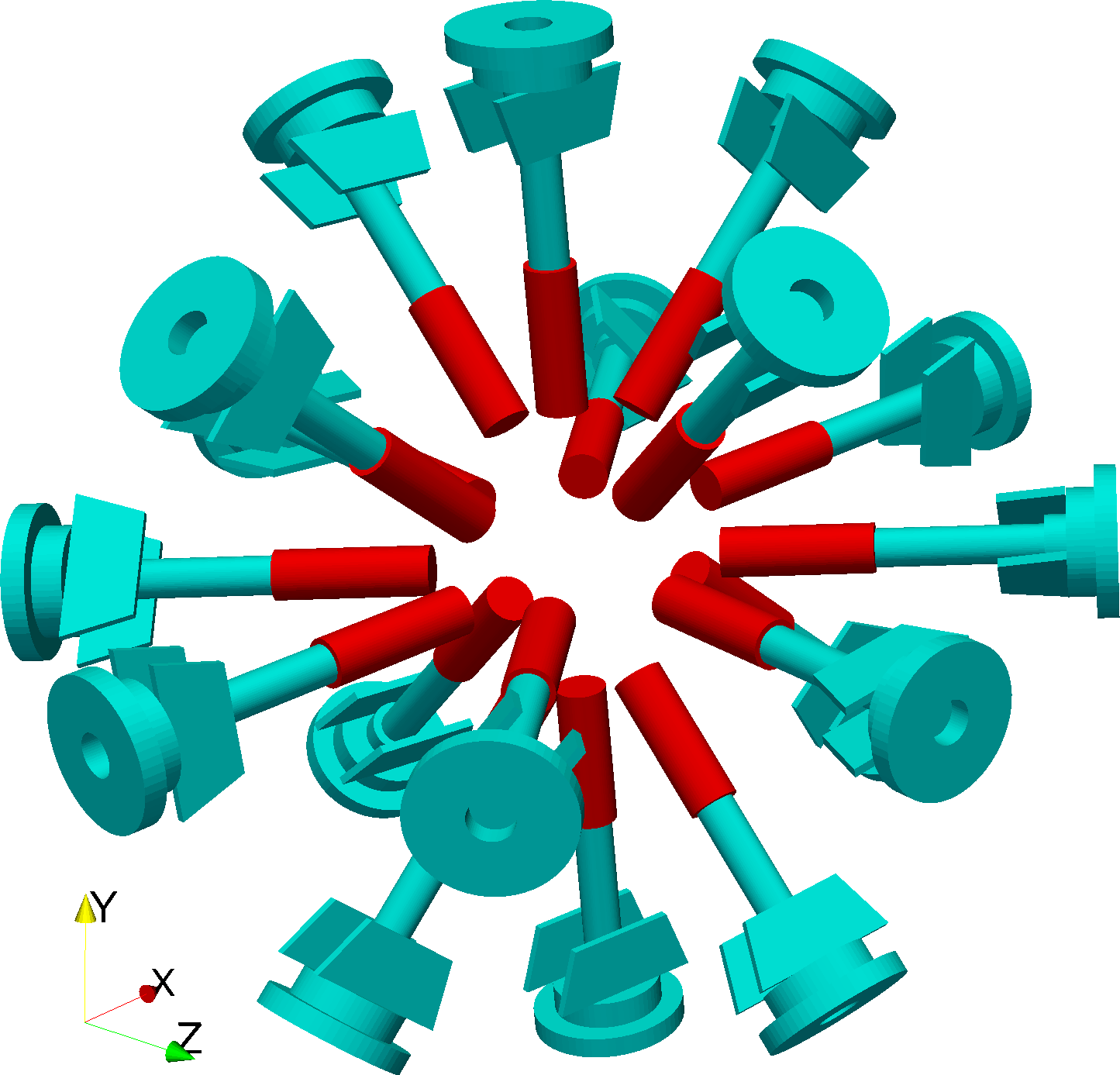}
		\caption{Tool 1.}
	\end{subfigure}%
	\begin{subfigure}[t]{0.33\linewidth}
		\centering
		\includegraphics[width=0.9\linewidth]{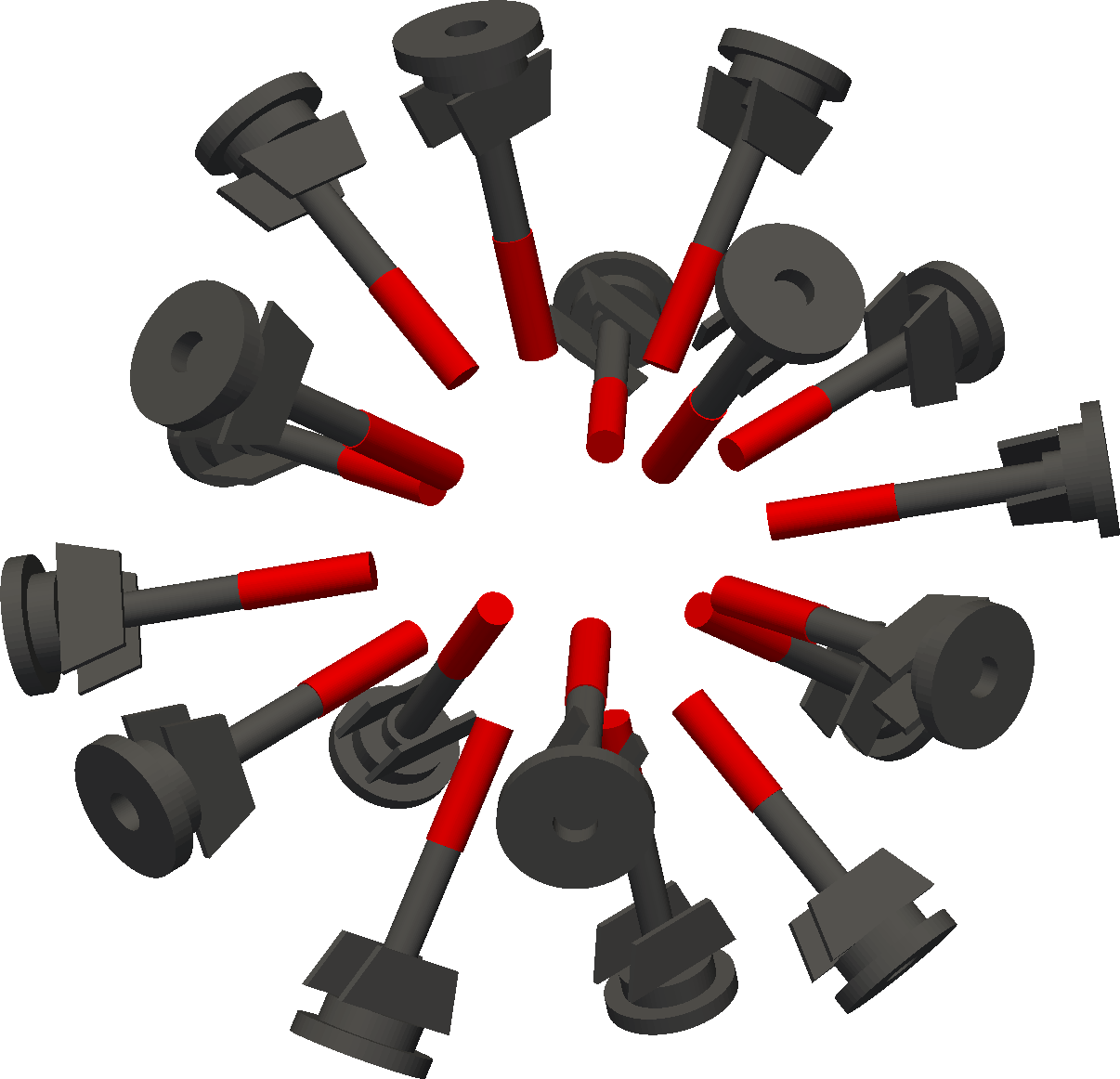}
		\caption{Tool 2.}
	\end{subfigure}%
	\begin{subfigure}[t]{0.33\linewidth}
		\centering
		\includegraphics[width=0.9\linewidth]{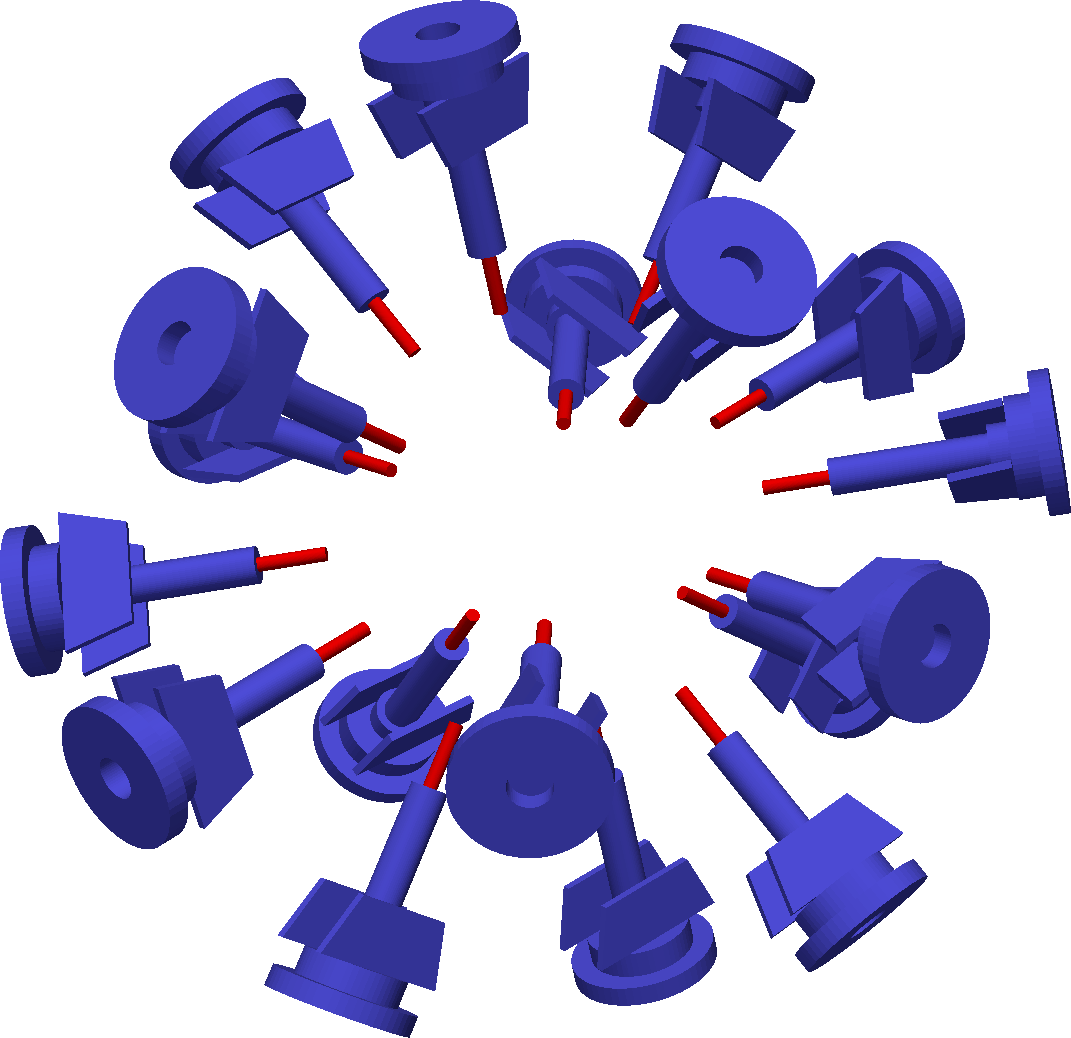}
		\caption{Tool 3.}
	\end{subfigure}%
	\caption{Three machining tools and their corresponding orientations.} \label{fig_tools}
\end{figure}

Next, let us consider the 3D example of Fig. \ref{fig_SuppBracketGeom}, which is a topologically optimized support bracket \cite{mirzendehdel2020topology}. We aim to find an optimized build orientation for the support bracket given the 4 fixturing configurations of Fig. \ref{fig_fixtures} and the 3 machining tools at different sizes shown in Fig. \ref{fig_tools}. Each tool has 18 orientations available, specifically given each tool is originally along $(0,1,0)$:
\begin{align}
\Theta= 	\{&R_{(1,0,0)}(0), R_{(1,0,0)}(\pi), R_{(1,0,0)}(\pi/2), R_{(1,0,0)}(3\pi/2),\nonumber\\
	&R_{(1,0,0)}(\pi/4),R_{(1,0,0)}(3\pi/4), R_{(1,0,0)}(5\pi/4),   \nonumber\\
	&R_{(1,0,0)}(7\pi/4),R_{(0,0,1)}(\pi/2), R_{(0,0,1)}(3\pi/2),\nonumber\\
	&R_{(0,0,1)}(\pi/4),R_{(0,0,1)}(3\pi/4), R_{(0,0,1)}(5\pi/4),\nonumber\\
	& R_{(0,0,1)}(7\pi/4),R_{(1,0,1)}(\pi/2), R_{(1,0,1)}(3\pi/2),\nonumber\\
	&R_{(1,0,-1)}(\pi/2),R_{(1,0,-1)}(3\pi/2)\nonumber\}
\end{align}
where $R_{\mathbf{a}}(\theta)$ denotes the rotation matrix in 3D by angle $\theta$ about axis $\mathbf{a}$.
The resolution is chosen such that there are about 100,000 voxels in the workpiece, i.e., the support bracket. For each cutter, we choose ten sharp points $\bk$ ($n_{\bK} = 10$).
For each case, we uniformly sample 100 build orientations ($n_\buildDir = 100$) to approximate $\SO{3}$ and select 5 build orientations ($n_{\buildDir^*} = 5$) with the lowest weighted Pareto-optimality value $\xi$. To identify secluded supports, we use $\lambda = 0.001$.
The reason behind providing multiple (here 5) optimized orientations is : 1) there may be other factors besides support volume and support accessibility that engineers need to consider but are not modeled in the optimization process and 2) to clearly demonstrate the impact of $w_{acc}$ as in some cases the “best” orientation is not affected by a slight change in $w_{acc}$, but the subsequent optimized orientations do change.
In general, since finding an optimized build orientation is a multi-objective optimization where not all aspects of the problem are modeled, it is better to provide multiple options to engineers to choose from. For all examples, $\volSuppMax = 1302.53$ (\cmm) and $\volSecMax = 47.61$ (\cmm).
 
\begin{figure} [h!]
	\centering
	\begin{subfigure}[t]{0.5\linewidth}
		\centering
		\includegraphics[width=0.99\linewidth]{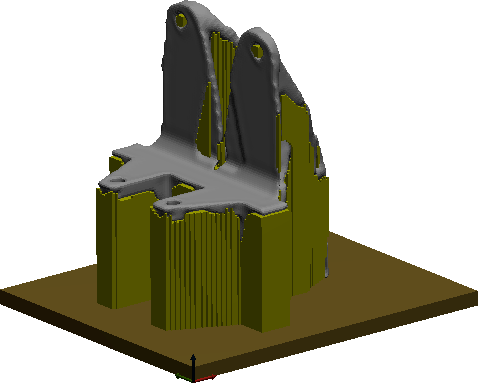}
		\caption{$\buildDir^*_1 = (0,0,1)$.}
	\end{subfigure}%
	\begin{subfigure}[t]{0.5\linewidth}
		\centering
		\includegraphics[width=0.99\linewidth]{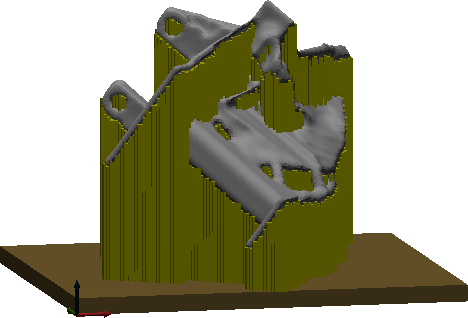}
		\caption{$\buildDir^*_2 = (-0.59,-0.81,-0.05)$.}
	\end{subfigure}%
	
	\begin{subfigure}[t]{0.5\linewidth}
		\centering
		\includegraphics[width=0.95\linewidth]{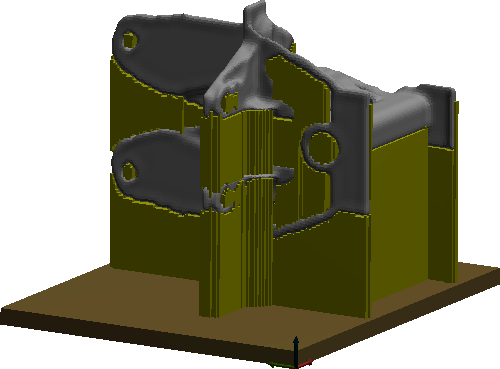}
		\caption{$\buildDir^*_3 = (0,-1,0)$.}
	\end{subfigure}%
	\begin{subfigure}[t]{0.5\linewidth}
		\centering
		\includegraphics[width=0.95\linewidth]{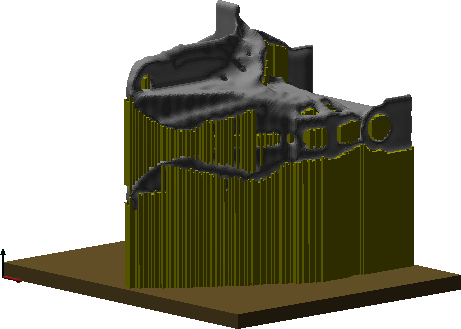}
		\caption{$\buildDir^*_4 = (0,1,0)$.}
	\end{subfigure}%
	
	\begin{subfigure}[t]{0.5\linewidth}
		\centering
		\includegraphics[width=0.99\linewidth]{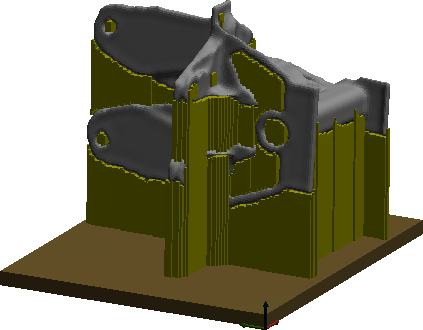}
		\caption{$\buildDir^*_5 = (-0.05,-1,0.03)$.}
	\end{subfigure}%
	\caption{5 Pareto-optimal directions with $w_{acc} = 0$.} \label{fig_nearnets_wt0}
\end{figure}

\begin{figure*} [h!]
	\begin{subfigure}[t]{0.75\linewidth}
		\centering
		\includegraphics[width=0.9\linewidth]{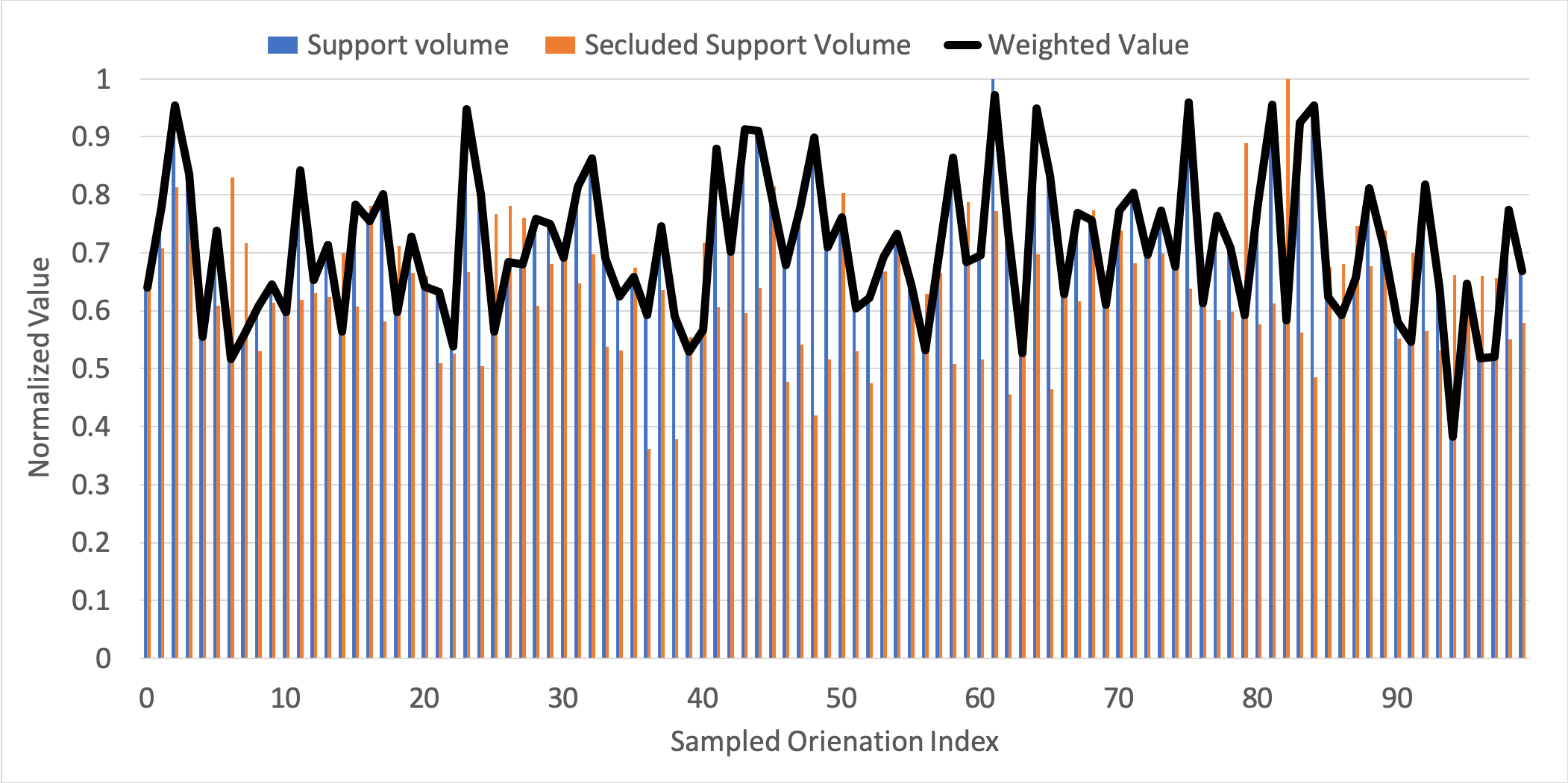}
		\caption{Sampled orientations and values.}
	\end{subfigure}%
	\begin{subfigure}[t]{0.25\linewidth}
		\centering
		\includegraphics[width=0.99\linewidth]{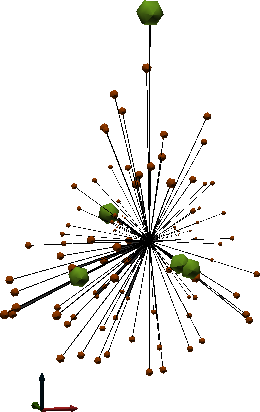}
		\caption{Sampled SO3.}
	\end{subfigure}%
	\caption {Sampled orientations and support volumes for $w_{acc} = 0$. The edge length and radius of each ball reflects the level of optimality within the sampled orientations. Selected build orientations are indicated in green. } \label{fig_SO3_wt0}
\end{figure*}

Figure \ref{fig_SO3_wt0}a shows the normalized total support volume, normalized secluded support volume, and the weighted values ($\{\xi_i\}$) indicating Pareto-optimality of each sampled build orientation $\buildDir_m$. Observe that since $w_{acc} = 0$, the weighted is the same as total support volume. Figure \ref{fig_SO3_wt0}b illustrates the sampled $\SO{3}$, where the edge length and radius of each ball reflects the level of optimality within the sampled orientations. Figure \ref{fig_nearnets_wt0} illustrates the near-net shapes for the 5 best build orientations w.r.t. total support volume. The best build orientation in this case is the canonical direction along +Z , where the support volume is 394 \cmm and secluded support volume is 31.49 \cmm. Values for other optimized solutions with $w_{acc} = 0$ are summarized in Table \ref{tab_wt0}.

\begin{table} [h!]
	\caption{Summary of results for $w_{acc} = 0$.}
	\tabulinesep=1mm
	\begin{tabu}[t!]{l|cccc}
	\hline \hline 
	& $\buildDir^*$ & $\volSupp$ (cm$^3$)  & $\volSec$ (cm$^3$) & $\xi$ \\
	\hline
	1&$(0,0,1)$ & 394 & 31.49 & 0.38\\
	2& $(-0.59,-0.81,-0.05)$ & 672 & 39.50 & 0.516 \\
	3&$(0,-1,0)$ & 674 & 31.44 &0.517 \\
	4&$(0,1,0)$ & 678 & 31.22 & 0.521 \\
	5&$(-0.05,-1,0.03)$ & 685 & 29.91 & 0.527\\
	\hline
\end{tabu}
	\label{tab_wt0}
\end{table}

\begin{figure*} [h!]
		\centering
	\begin{subfigure}[t]{0.75\linewidth}
		\centering
		\includegraphics[width=0.9\linewidth]{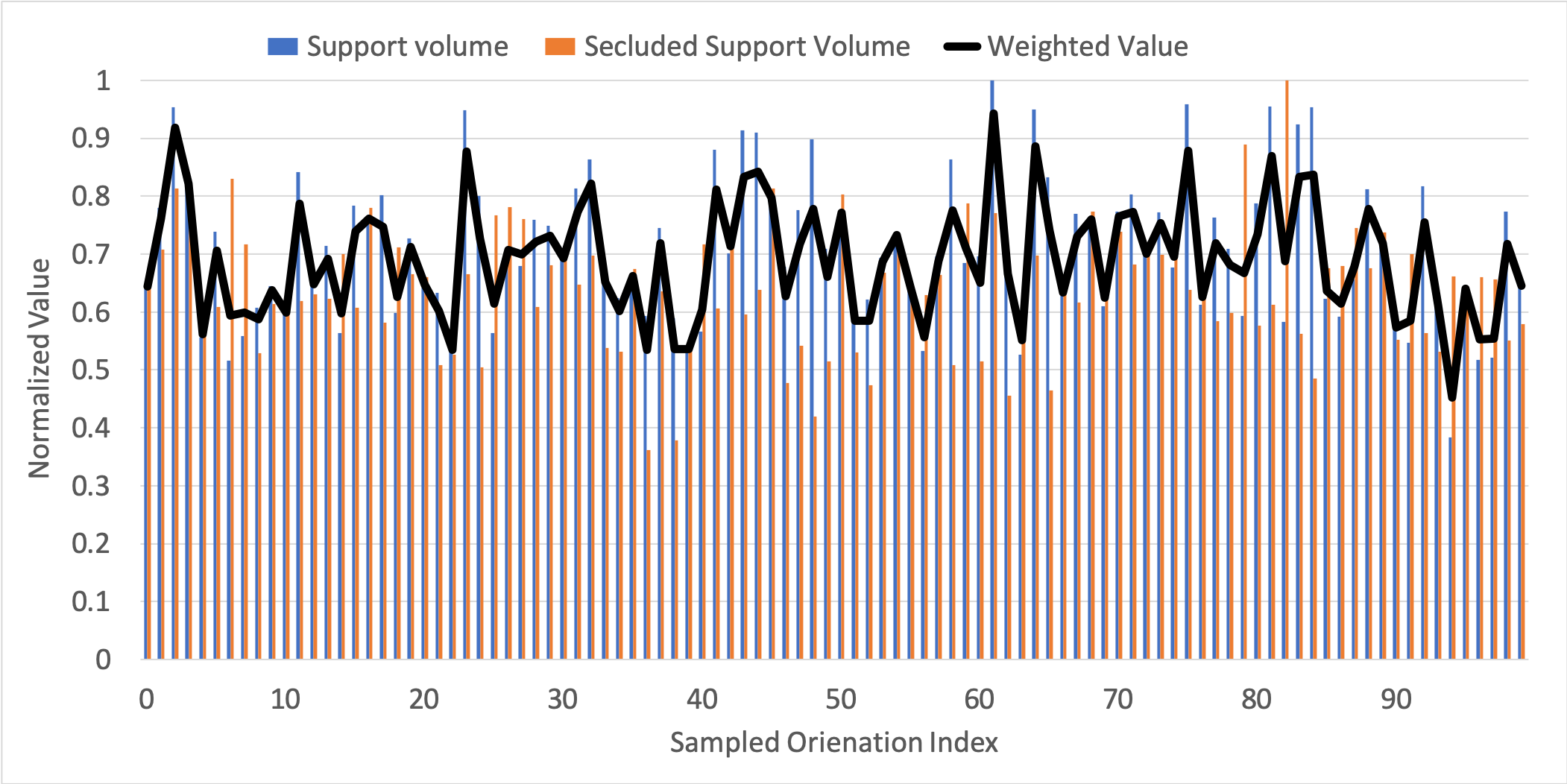}
		\caption{Sampled orientations and values.}
	\end{subfigure}%
	\begin{subfigure}[t]{0.25\linewidth}
		\centering
		\includegraphics[width=0.99\linewidth]{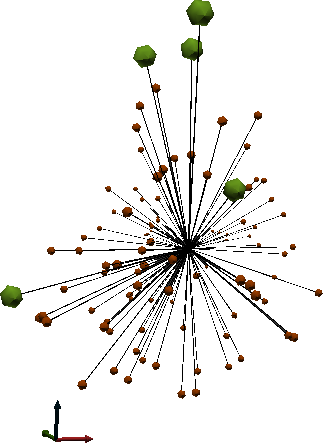}
		\caption{Sampled SO3.}
	\end{subfigure}%
	\caption{Sampled orientations and support volumes for $w = 0.25$.} \label{fig_SO3_wt25}
\end{figure*}

Figure \ref{fig_SO3_wt25}a shows the normalized support volumes and the weighted Pareto-optimality values for $w_{acc} = 0.25$. In this case, the optimality of each orientation mostly relies on support volume and removability is of secondary concern. Figure \ref{fig_SO3_wt25}b illustrates the sampled $\SO{3}$ for $w_{acc} = 0.25$.

\begin{figure} [h!]
	\centering
	\begin{subfigure}[t]{0.5\linewidth}
		\centering
		\includegraphics[width=0.99\linewidth]{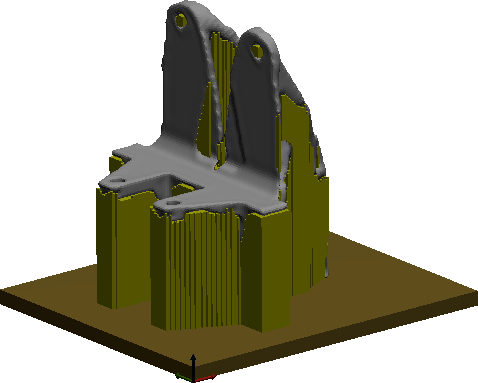}
		\caption{$\buildDir^*_1 = (0,0,1)$}
	\end{subfigure}%
	\begin{subfigure}[t]{0.5\linewidth}
		\centering
		\includegraphics[width=0.99\linewidth]{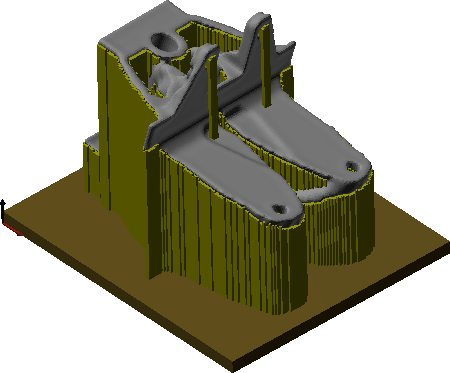}
		\caption{$\buildDir^*_2 = (-0.97,-0.13,-0.20)$}
	\end{subfigure}%
	
	\begin{subfigure}[t]{0.5\linewidth}
		\centering
		\includegraphics[width=0.9\linewidth]{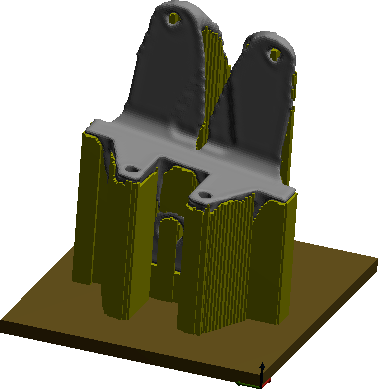}
		\caption{$\buildDir^*_3 = (0.01,0.14,0.99)$}
	\end{subfigure}%
	\begin{subfigure}[t]{0.5\linewidth}
		\centering
		\includegraphics[width=0.99\linewidth]{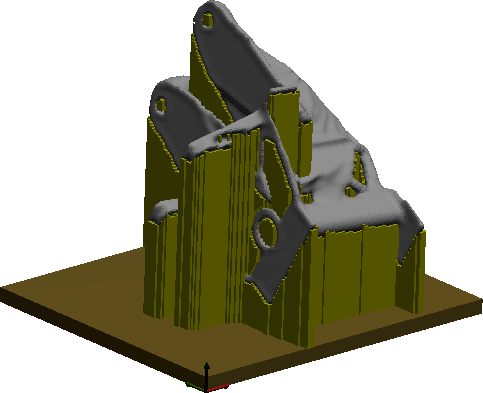}
		\caption{$\buildDir^*_4 = (-0.08,-0.88,0.47)$}
	\end{subfigure}%
	
	\begin{subfigure}[t]{0.5\linewidth}
		\centering
		\includegraphics[width=0.7\linewidth]{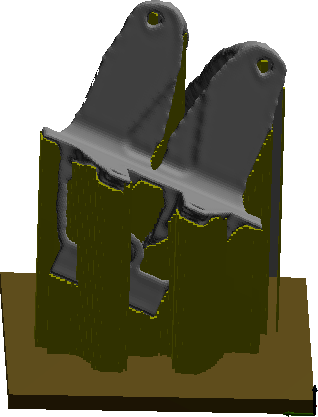}
		\caption{$\buildDir^*_5 = (-0.15,0.38,-0.91)$}
	\end{subfigure}%
	\caption{ 5 Pareto-optimal directions with $w = 0.25$.} \label{fig_nearnets_wt25}
\end{figure}

Figure \ref{fig_nearnets_wt25} illustrates the near-net shapes for the 5 best build orientations w.r.t. the weighted Pareto-optimality criteria. Similar to the previous case, the first optimized orientation is (0,0,1), however the $\xi$ has increased from 0.38 to 0.45; in other words, its level of Pareto-optimality has reduced. Further, according to Table \ref{tab_wt25}, other orientations have changed, indicating that the two objectives are in many cases competing.

\begin{table} [h!]
	\caption{Summary of results for $w_{acc} = 0.25$.}
	\tabulinesep=1mm
	\begin{tabu}[t!]{l|cccc}
	\hline \hline 
	& $\buildDir^*$ & $\volSupp$ (cm$^3$)  & $\volSec$ (cm$^3$) & $\xi$ \\
	\hline
	1&$(0,0,1)$ & 394 & 31.49 & 0.453\\
	2& $(-0.97,-0.13,-0.20)$ & 772 & 17.22 & 0.5352 \\
	3&$(0.01,0.14,0.99)$ & 701 & 25.05&0.5353 \\
	4&$(-0.08,-0.88,0.47)$ & 690 & 26.37 & 0.5357 \\
	5&$(-0.15,0.38,-0.91)$ & 767 & 17.99 & 0.5364\\
	\hline
\end{tabu}
	\label{tab_wt25}
\end{table}

\begin{figure} [h!]
	\centering
	\begin{subfigure}[t]{0.5\linewidth}
		\centering
		\includegraphics[width=0.99\linewidth]{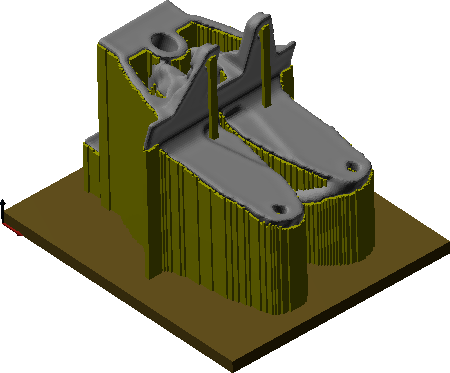}
		\caption{$\buildDir^*_1 = (-0.97,-0.13,-0.20)$}
	\end{subfigure}%
	\begin{subfigure}[t]{0.5\linewidth}
		\centering
		\includegraphics[width=0.73\linewidth]{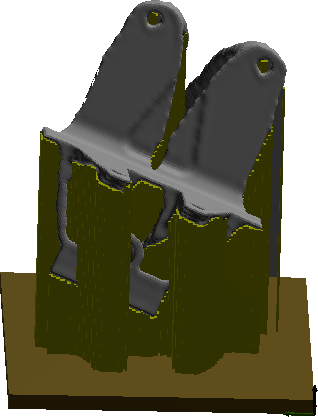}
		\caption{$\buildDir^*_2 = (-0.15,0.38,-0.91)$}
	\end{subfigure}%
	
	\begin{subfigure}[t]{0.55\linewidth}
		\centering
		\includegraphics[width=0.99\linewidth]{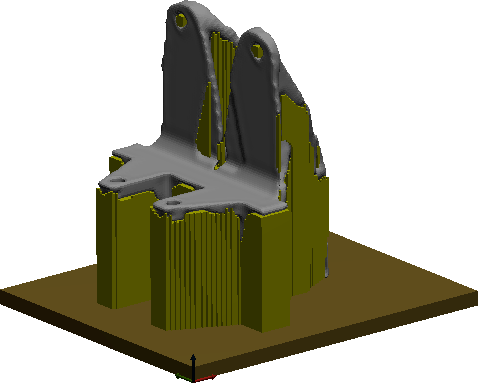}
		\caption{$\buildDir^*_3 = (0,0,1)$}
	\end{subfigure}%
	\begin{subfigure}[t]{0.45\linewidth}
		\centering
		\includegraphics[width=0.99\linewidth]{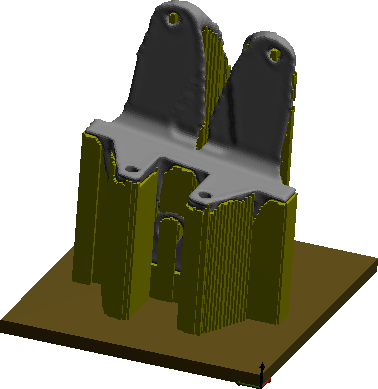}
		\caption{$\buildDir^*_4 = (0.01,0.14,0.99)$}
	\end{subfigure}%
	
	\begin{subfigure}[t]{0.5\linewidth}
		\centering
		\includegraphics[width=0.99\linewidth]{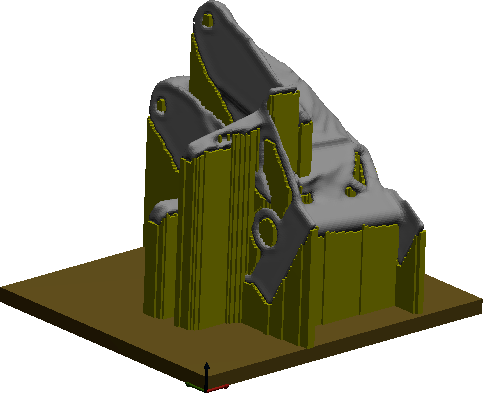}
		\caption{$\buildDir^*_5 = (-0.08,-0.88,0.47)$}
	\end{subfigure}%
	\caption{5 Pareto-optimal directions with $w = 0.50$.} \label{fig_nearnets_wt50}
\end{figure}

\begin{figure*} [h!]
		\centering
	\begin{subfigure}[t]{0.75\linewidth}
		\centering
		\includegraphics[width=0.9\linewidth]{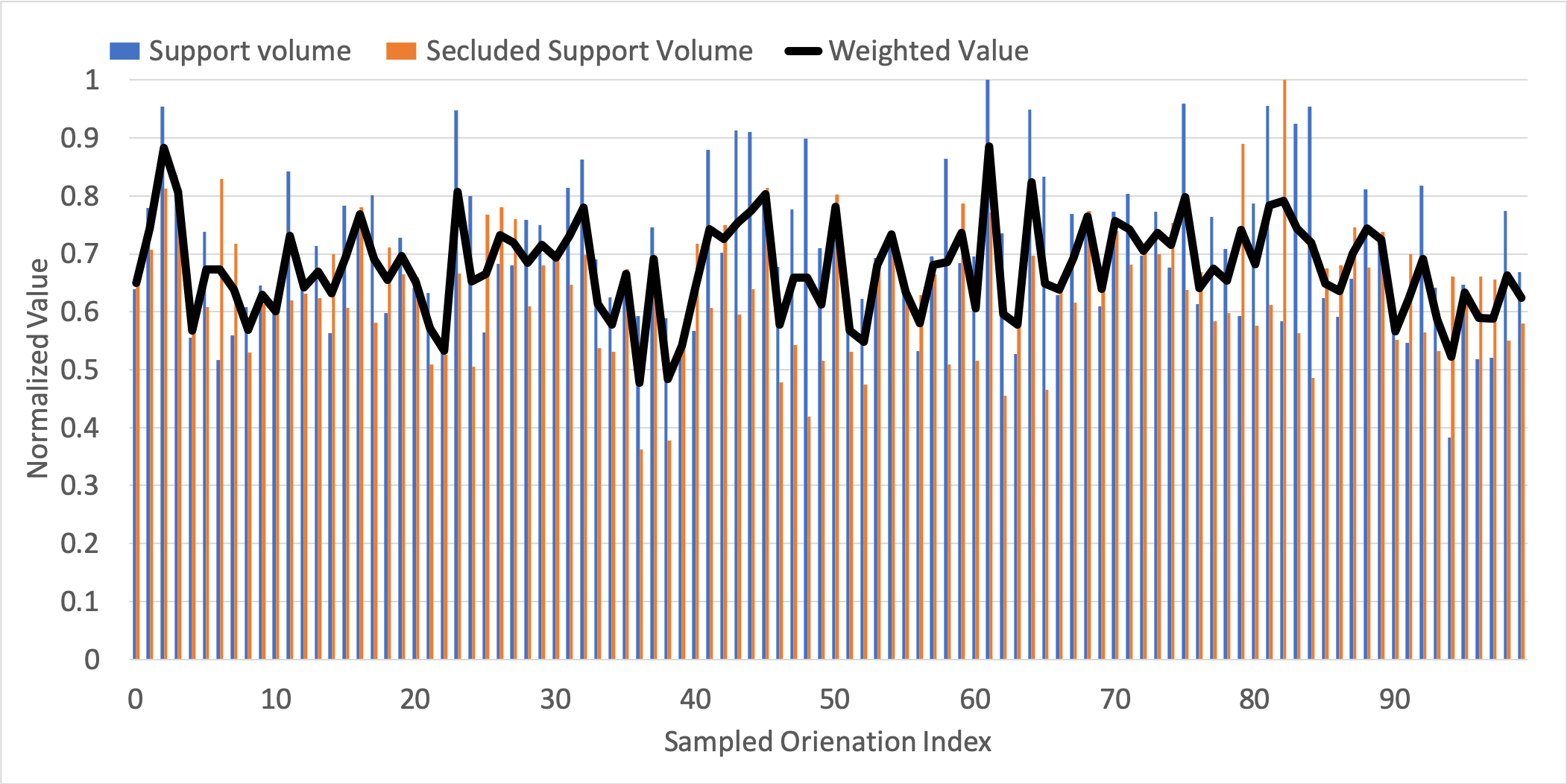}
		\caption{Sampled orientations and values.}
	\end{subfigure}%
	\begin{subfigure}[t]{0.25\linewidth}
		\centering
		\includegraphics[width=0.99\linewidth]{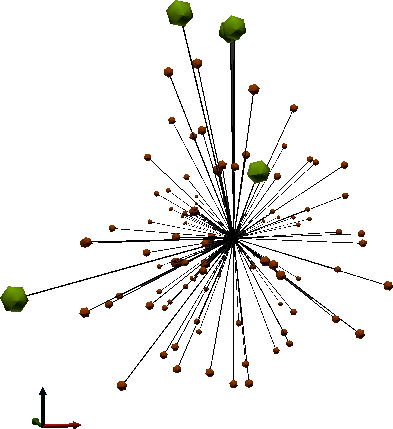}
		\caption{Sampled SO3.}
	\end{subfigure}%
	\caption{Sampled orientations and support volumes for $w = 0.50$.} \label{fig_SO3_wt50}
\end{figure*}

Figure \ref{fig_SO3_wt50} shows the normalized support volumes and $\{\xi_i\}$ over the sampled $\SO{3}$ for $w_{acc} = 0.5$. Figure \ref{fig_nearnets_wt50} illustrates the selected optimized build orientations. The first optimized is $\buildDir_1 =(-0.97,-0.13,-0.20)$ with support volume of 772 \cmm and secluded volume of 17.22 \cmm. Observe that compared to the previous cases, total support volume is significantly higher, however the reduction in secluded volume has made it the Pareto-optimal orientation for $w_{acc} = 0.5$. The values for other selected build orientations are presented in Table \ref{tab_wt50}. 

\begin{table} [h!]
	\caption{Summary of results for $w_{acc} = 0.50$.}
	\tabulinesep=1mm
	\begin{tabu}[t!]{l|cccc}
		\hline \hline 
		 & $\buildDir^*$ & $\volSupp$ (cm$^3$)  & $\volSec$ (cm$^3$) & $\xi$ \\
		\hline
		1& $(-0.97,-0.13,-0.20)$ & 772 & 17.22 & 0.477 \\
		2&$(-0.15,0.38,-0.91)$ & 767 & 17.99 & 0.483\\
		3&$(0,0,1)$ & 394 & 31.49 & 0.522\\
		4&$(0.01,0.14,0.99)$ & 701 & 25.05&0.532 \\
		5&$(-0.08,-0.88,0.47)$ & 690 & 26.37 & 0.542 \\
		\hline
	\end{tabu}
	\label{tab_wt50}
\end{table}

\begin{figure} [h!]
	\centering
	\begin{subfigure}[t]{0.5\linewidth}
		\centering
		\includegraphics[width=0.99\linewidth]{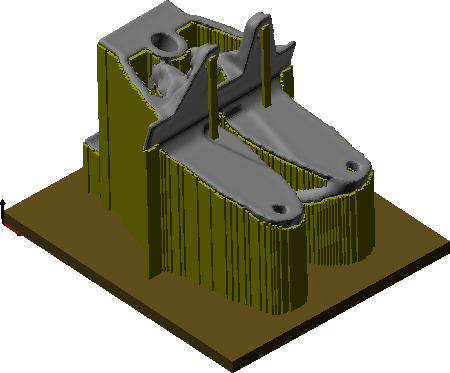}
		\caption{$\buildDir^*_1 = (-0.97,-0.13,-0.20)$}
	\end{subfigure}%
	\begin{subfigure}[t]{0.5\linewidth}
		\centering
		\includegraphics[width=0.75\linewidth]{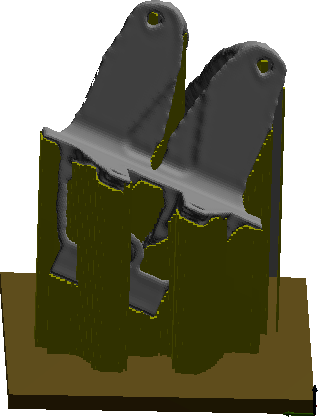}
		\caption{$\buildDir^*_2 = (-0.15,0.39,0.91)$.}
	\end{subfigure}%
	
	\begin{subfigure}[t]{0.5\linewidth}
		\centering
		\includegraphics[width=0.99\linewidth]{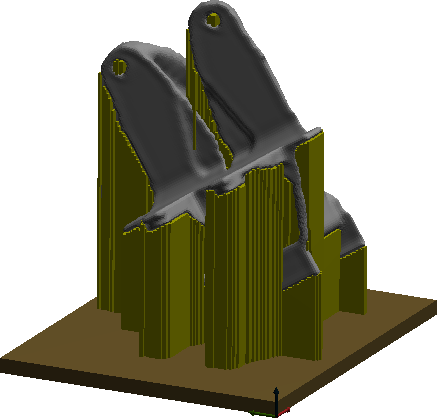}
		\caption{$\buildDir^*_3 = (-0.02,-0.53,0.84)$}
	\end{subfigure}%
	\begin{subfigure}[t]{0.5\linewidth}
		\centering
		\includegraphics[width=0.99\linewidth]{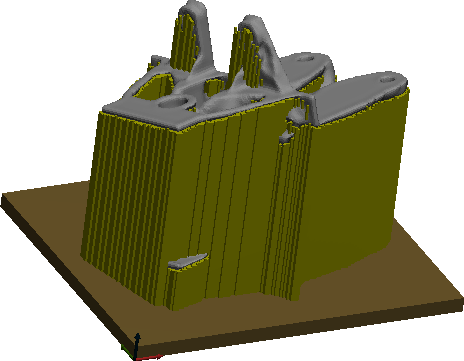}
		\caption{$\buildDir^*_4 = (-0.95,-0.29,0.06)$}
	\end{subfigure}%
	
	\begin{subfigure}[t]{0.5\linewidth}
		\centering
		\includegraphics[width=0.99\linewidth]{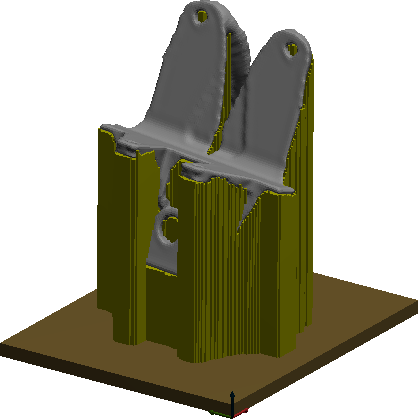}
		\caption{$\buildDir^*_5 = (-0.35,0.08,0.93)$}
	\end{subfigure}%
	\caption{ 5 Pareto-optimal directions with $w = 0.75$.} \label{fig_nearnets_wt75}
\end{figure}

\begin{figure*} [h!]
		\centering
	\begin{subfigure}[t]{0.75\linewidth}
		\centering
		\includegraphics[width=0.9\linewidth]{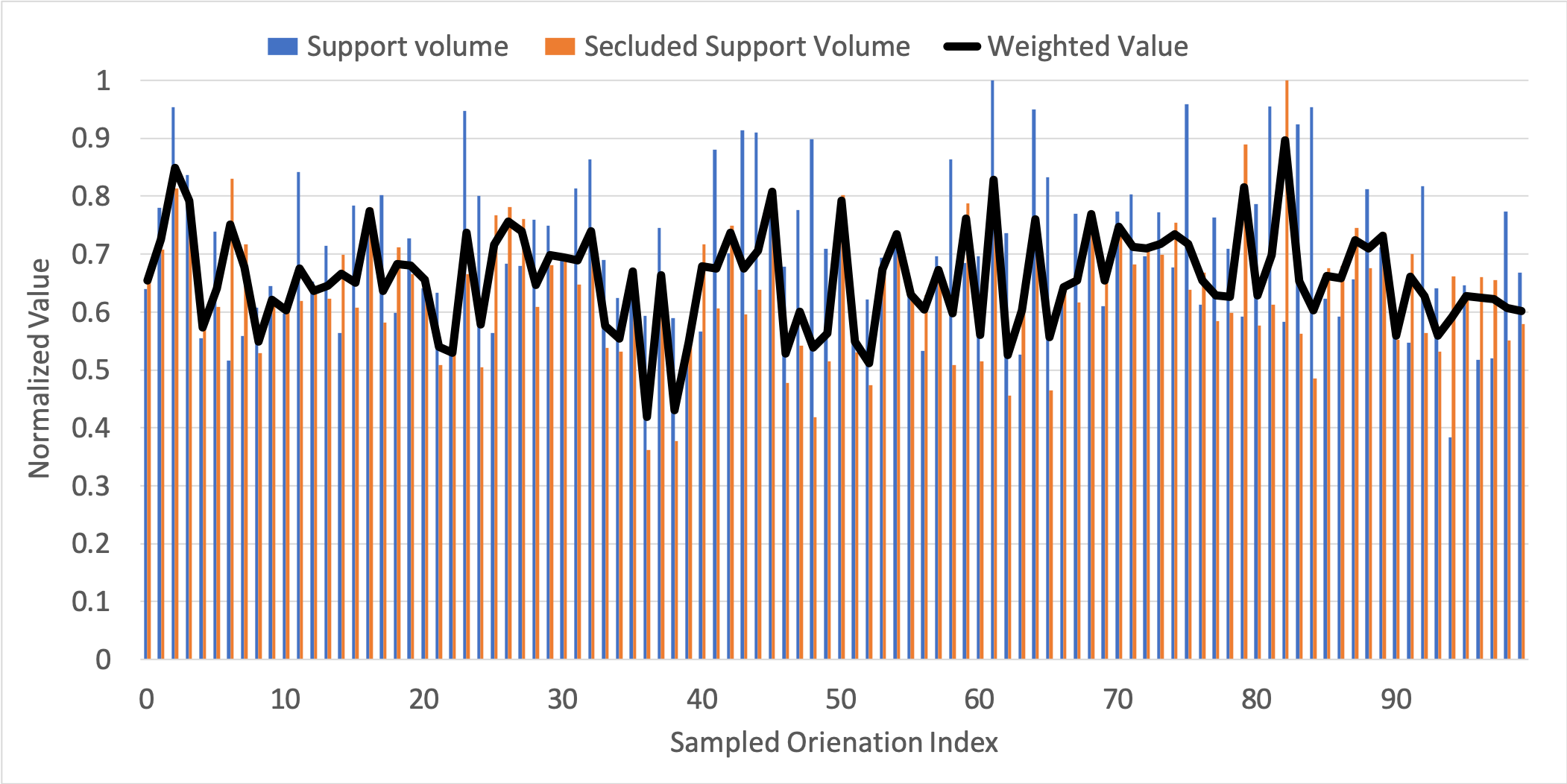}
		\caption{Sampled orientations and values.}
	\end{subfigure}%
	\begin{subfigure}[t]{0.25\linewidth}
		\centering
		\includegraphics[width=0.99\linewidth]{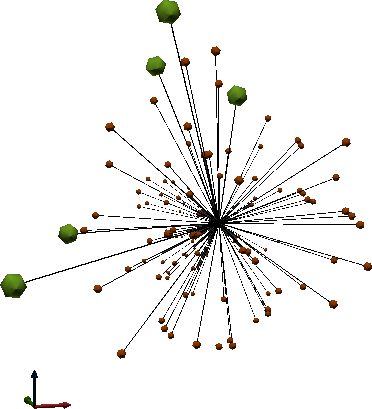}
		\caption{Sampled SO3.}
	\end{subfigure}%
	\caption{Sampled orientations and support volumes for $w = 0.75$.} \label{fig_SO3_wt75}
\end{figure*}

As shown in Fig. \ref{fig_nearnets_wt75} and indicated in Table \ref{tab_wt75}, the optimized build orientation for $w_{acc} = 0.75$ is similar to the one obtained for $w_{acc} = 0.50$ with $\buildDir_1 =(-0.97,-0.13,-0.20)$. In this case, since the optimality is mainly due to low secluded volume, increasing $w_{acc}$ results in a decrease in $\xi$ from 0.47 to 0.42. Figure \ref{fig_SO3_wt75} illustrates the sampled orientation space and values for quantities of interest.

\begin{table} [h!]
	\caption{Summary of results for $w_{acc} = 0.75$.}
	\tabulinesep=1mm
	\begin{tabu}[t!]{l|cccc}
	\hline \hline 
	& $\buildDir^*$ & $\volSupp$ (cm$^3$)  & $\volSec$ (cm$^3$) & $\xi$ \\
	\hline
	1& $(-0.97,-0.13,-0.20)$ & 772 & 17.22 & 0.42 \\
	2&$(-0.15,0.38,-0.91)$ & 767 & 17.99 & 0.43\\
	3&$(-0.02,-0.53,0.84)$ & 810 & 22.57 & 0.51\\
	4&$(-0.95,-0.29,0.06)$ & 958 & 21.68&0.52 \\
	5&$(-0.35,0.08,0.93)$ & 883 & 22.75 & 0.53 \\
	\hline
	\end{tabu}
	\label{tab_wt75}
\end{table}

\begin{figure} [!ht]
	\centering
	\begin{subfigure}[t]{0.5\linewidth}
		\centering
		\includegraphics[width=0.99\linewidth]{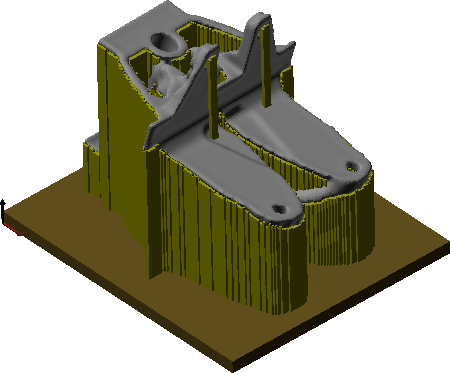}
		\caption{$\buildDir^*_1 = (-0.97,-0.13,-0.20)$}
	\end{subfigure}%
	\begin{subfigure}[t]{0.5\linewidth}
		\centering
		\includegraphics[width=0.75\linewidth]{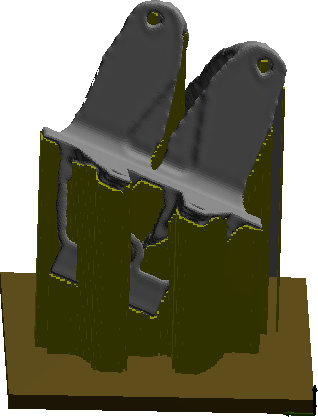}
		\caption{$\buildDir^*_2 =(-0.15,0.38,-0.91)$}
	\end{subfigure}%

	\begin{subfigure}[t]{0.5\linewidth}
		\centering
		\includegraphics[width=0.9\linewidth]{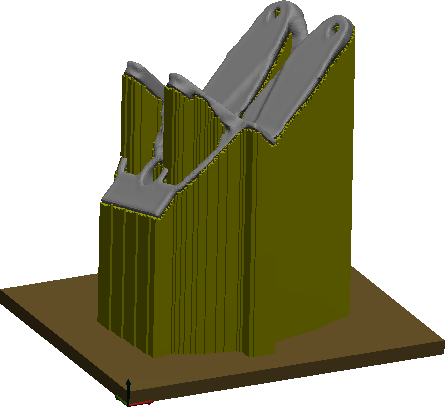}
		\caption{$\buildDir^*_3 = (-0.71,-0.20,0.67)$.}
	\end{subfigure}%
	\begin{subfigure}[t]{0.5\linewidth}
		\centering
		\includegraphics[width=0.99\linewidth]{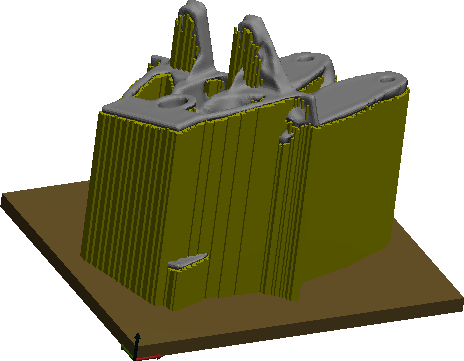}
		\caption{$\buildDir^*_4 = (-0.95,-0.29,0.05)$.}
	\end{subfigure}%

	\begin{subfigure}[t]{0.5\linewidth}
		\centering
		\includegraphics[width=0.99\linewidth]{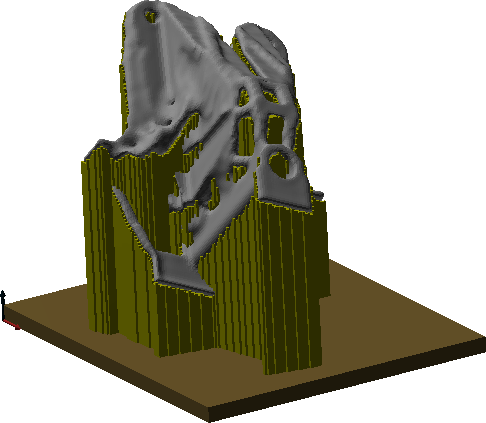}
		\caption{$\buildDir^*_5 = (0.73,-0.38,0.57)$.}
	\end{subfigure}%
	\caption{5 Pareto-optimal directions with $w =1.0$.} \label{fig_nearnets_wt100}
\end{figure}

\begin{figure*} [h!]
		\centering
	\begin{subfigure}[t]{0.75\linewidth}
		\centering
		\includegraphics[width=0.9\linewidth]{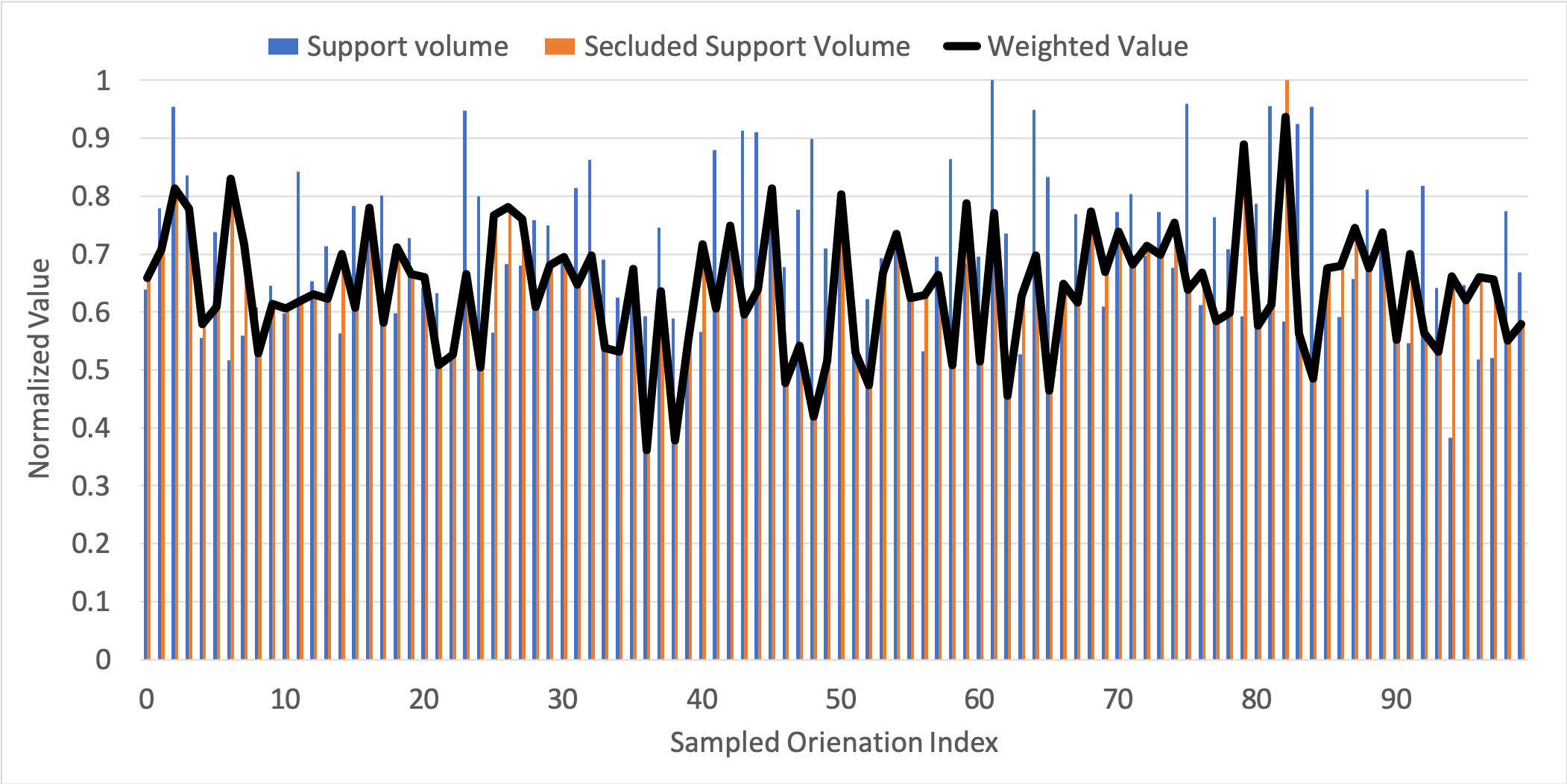}
		\caption{Sampled orientations and values.}
	\end{subfigure}%
	\begin{subfigure}[t]{0.25\linewidth}
		\centering
		\includegraphics[width=0.99\linewidth]{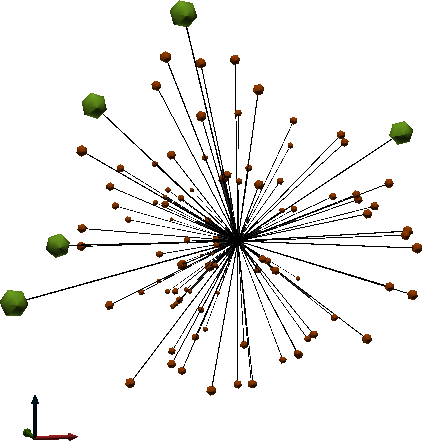}
		\caption{Sampled SO3.}
	\end{subfigure}%
	\caption{Sampled orientations and support volumes for $w = 1.0$.} \label{fig_SO3_wt100}
\end{figure*}

Finally, by increasing the value of $w_{acc}$ to 1.0 we are essentially ignoring support volume and only optimizing w.r.t. removability of the sacrificial support structure as illustrated in Fig. \ref{fig_SO3_wt100}. Compared to the previous case the first optimized build orientation has remained the same, however the value of $\xi$ has reduced to 0.36 and the other orientations have either swapped places or have been replaced as shown in Figure \ref{fig_nearnets_wt100}. Table \ref{tab_wt100} summarizes the results $w_{acc} = 1$. 

\begin{table} [h!]
	\caption{Summary of results for $w_{acc} = 1.0$.}
	\tabulinesep=1mm
	\begin{tabu}[t!]{l|cccc}
	\hline \hline 
	& $\buildDir^*$ & $\volSupp$ (cm$^3$)  & $\volSec$ (cm$^3$) & $\xi$ \\
	\hline
	1& $(-0.97,-0.13,-0.20)$ & 772 & 17.22 & 0.36 \\
	2&$(-0.15,0.38,-0.91)$ & 767 & 17.99 & 0.38\\
	3&$(-0.71,-0.20,0.67)$ & 1170 & 19.95 & 0.42\\
	4&$(-0.95,-0.29,0.05)$ & 958 & 21.68&0.45 \\
	5&$(0.73,-0.38,0.57)$ & 1085 & 22.12 & 0.46 \\
	\hline
\end{tabu}
	\label{tab_wt100}
\end{table}


\section{Support Removal Planning} \label{sec_planning}

In this section, we describe an automated support removal plan based on the IMF described in Section \ref{sec_imf}. The underlying assumption is that the entirety of the support region (of arbitrary shape) need to be machined. This is in contrast with a previous approach presented in \cite{nelaturi2019automatic}, where we made restrictive assumptions on the shape of the support structure; namely, a finite collection of support beams (e.g., vertical beams) attached at small `dislocation features' modeled as singular attachment points. The main drawback of this approach is the restrictive assumptions on both tool and support column shapes, and the inability to cut through the middle of columns to reach deeper columns, leading to unnecessary spirals around the part to remove support in ``rings'' of support columns.

The proposed greedy algorithm finds a sequence of actions in terms of fixturing, cutting tool and its orientation based on maximal removable volume. As described in Algorithm \ref{alg_suppRemPlan}, at each step the maximum accessible support volume for a single oriented tool and fixturing configuration is determined as explained in Algorithm \ref{alg_maxRem}. Subsequently, the near-net shape is updated and the process is repeated until the entirety of accessible support regions are removed. In practice, we halt the planning once the maximum removed volume is less than some threshold, here 0.5\% of support volume. 
 
 \begin{algorithm} [ht!]
 	\caption{Support Removal Planning.}
 	\begin{algorithmic}
 		\Procedure{Plan}{$\indic_\Omega,\buildDir $}
 		
 		\State $\indic_{\Omega_\buildDir} \gets \Call{Rotate}{\indic_\Omega,\buildDir}$ 
 		\State $\indic_{S^{(0)}} \gets \Call{SuppGen}{\indic_{\Omega_\buildDir} }$ \Comment{Generate support}
 		\State $\volSupp \gets \Call{Volume}{\indic_{S^{(0)}}}$ \Comment{Compute support volume}
		\State $\indic_{N^{(0)}} \gets \indic_{\Omega_\buildDir} + \indic_{S^{(0)}}$ \Comment{Near-net shape}
		\State $ s \gets 1$ \Comment{Initialize planning step}
 		\While{$\volSupp > \epsilon$ } 
 		\State  $(\indic_{S_\text{rem}^{(s)}},\cdots)\gets\Call{Remove}{\indic_{\Omega_\buildDir},\cdots}$ 	
 		\State $\indic_{S^{(s)}} \gets  \indic_{S_\text{rem}^{(s)}} - \indic_{S^{(s-1)}}$
 		\State $\indic_{N^{(s)}} \gets \indic_{\Omega_\buildDir} +  \indic_{S^{(s)}}$ \Comment{Implicit difference}
 		\State $\volSupp \gets \Call{Volume}{\indic_{S^{(s)}}}$ \Comment{Update Near-net shape}
 		\State $ s \gets s+1$ \Comment{Next step}
		\EndWhile
 		\State \Return{($\indic_{N^{(s)}},\indic_{T^{(s)}},\indic_{F^{(s)}}$)}
 		\EndProcedure 
 	\end{algorithmic} \label{alg_suppRemPlan}
 \end{algorithm}
 
\begin{algorithm} [ht!]
	\caption{Find Maximum Removable Support.}
	\begin{algorithmic}
		\Procedure{Remove}{$[\indic_\Omega], [\indic_S],[\indic_P], [\indic_{F}], [\indic_{H}],[\indic_{K}], \{\Theta\}$)}
		\State Initialize $V_\text{max} \gets 0$
		\For{$j \gets 1$ to $n_F$}
		\State Define $[\indic_{O_j}] \gets [\indic_{\Omega}] + [\indic_{F_j}] + [\indic_P]$	\Comment{Union}
		\State Initialize $[f_\text{IMF}^{(j)}] \gets 0$ 
		\For{$i \gets 1$ to $n_T$}
		\State Define $[\indic_{T_i}] \gets [\indic_{H_i}] + [\indic_{K_i}]$
		\Comment{Implicit union}
		\State Initialize $[\gamma_i] \gets [0]$ 
		\Comment{IMF for the $i^\text{th}$ tool}
		\ForAll{$R \in \Theta_i$}
		\State $[\indic_{RT_i}] \gets \Call{Rotate}{[\indic_{T_i}], R}$
		\State $[\tilde{\indic}_{RT_i}] \gets \Call{Reflect}{[\indic_{RT_i}]}$
		\State $[g] \gets \Call{Convolve}{[\indic_{O_j}], [\tilde{\indic}_{R T_i}]}$
		\ForAll{$\bk \in \Call{Support}{[\indic_{K_i}]}$}
		\State $[h] \gets \Call{Translate}{[g], -R \bk}$
		\State $[\gamma_i] \gets \min([\gamma_i], [h])$
		\EndFor
		\State $\indic_{S_\text{rem}} \gets [\indic_S]\cdot[\gamma_i > \lambda]$
		\State $V_\text{rem} \gets \Call{Volume}{\indic_{S_\text{rem}}}$
		\If{$ V_\text{rem} > V_\text{max}$}
		\State $V_\text{max} \gets V_\text{rem}$
		\State $ \indic_{S_\text{max}} \gets \indic_{S_\text{rem}}$, 
		 $ \indic_{T_\text{max}} \gets \indic_{T_i}$
		 \State $ \indic_{R_\text{max}} \gets \indic_R $, $ \indic_{F_\text{max}} \gets \indic_{F_j}$
		\EndIf	
		\EndFor
		\EndFor
		\EndFor
		\State\Return{($\indic_{S_\text{max}}, \indic_{T_\text{max}},\indic_{R_\text{max}},\indic_{F_\text{max}}$)}
		\EndProcedure 
	\end{algorithmic} \label{alg_maxRem}
\end{algorithm}
 
\begin{figure*} [ht!]
 	\centering
 	\includegraphics[width=0.8\linewidth]{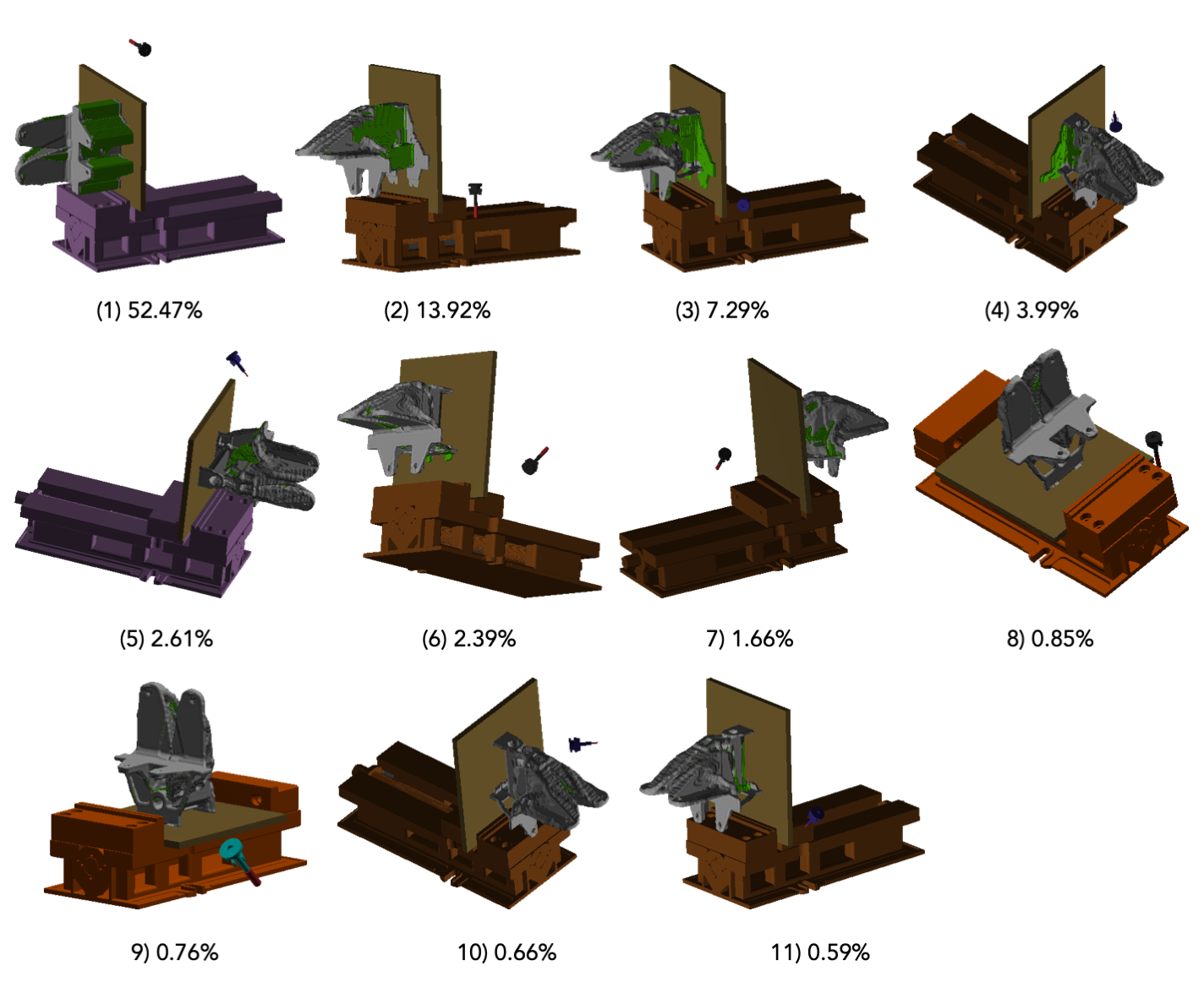}
 	\caption{11-step support removal plan for an optimized orientation with $w_{acc} = 0$ where about 87 \% of supports can be removed.} \label{fig_plan_wt0_opt1}
 \end{figure*}

 Figure \ref{fig_plan_wt0_opt1} illustrates an 11-step support removal plan for the optimized build orientation $\buildDir^*_1 = (0,0,1)$ obtained for $w_{acc} = 0$ and $w_{acc} = 0.25$. Table \ref{tab_planWt0} summarizes the removed volume fraction at each step as well as fixturing device, tool and its approach direction. Overall, 87\% of the supports are removed.   
 
 \begin{table} [h!]
 	\caption{Summary of results for plan for the optimized build orientation  $\buildDir^*_1 = (0,0,1)$ at $w_{acc} = 0$ with 93\% removable supports.}
 	\tabulinesep=1mm
 	\begin{tabu}[t!]{lcccl}
 		\hline \hline 
 		Step &Vol. Frac. (\%) & Fixture  & Tool& Direction\\
 		\hline
 		1 & 52.47 & 4 & 2 & (1,0,0)\\
		2 & 13.92& 3 & 2 & (-1,0,0)\\
		3 & 7.29 & 3 & 3 & (0,1,-1)\\
		4 & 3.99 & 3 & 3 & (0,-1,-1)\\
		5 & 2.61 & 4 & 3 & (0,1,1)\\
		6 & 2.39 & 3 & 2& (1,0,-1)\\
		7 & 1.66 & 3 & 2& (-1,-1,0)\\
		8 & 0.85 & 1 & 2& (0,0,-1)\\
		9 & 0.76 & 1 & 1& (1,0,-1)\\
		10 & 0.66 & 3 & 3& (0,-1,1)\\
		11 & 0.59 & 3 & 3& (-1,1,0)\\
 		\hline
 	\end{tabu}
 	\label{tab_planWt0}
 \end{table}

\begin{figure*} [ht!]
	\centering
	\includegraphics[width=0.8\linewidth]{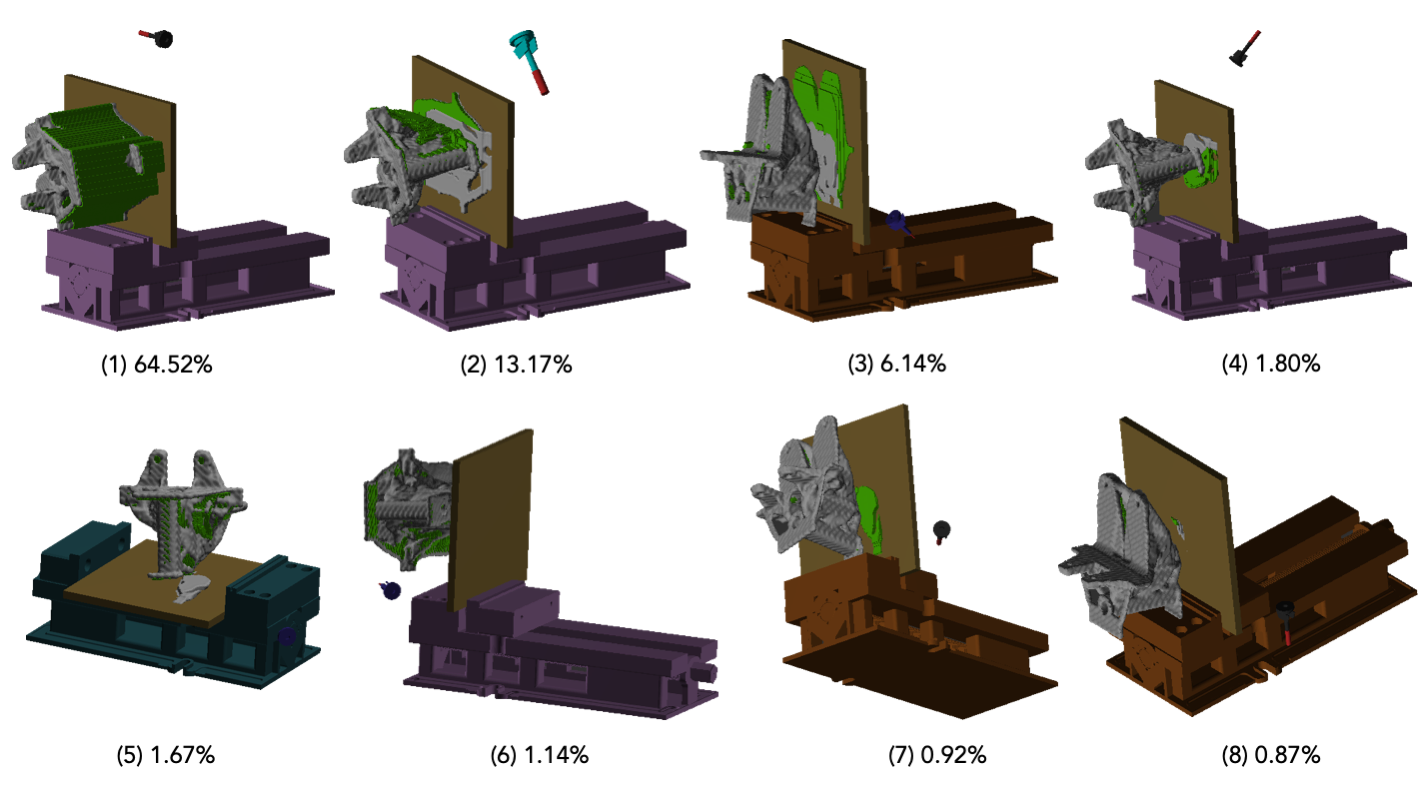}
	\caption{8-step support removal plan for an optimized orientation with $w_{acc} = 0.5$ where about 90 \% of supports can be removed.} \label{fig_plan_wt50_opt1}
\end{figure*}

Figure \ref{fig_plan_wt50_opt1} illustrates an 8-step support removal plan for the optimized build orientation $\buildDir^*_1 = (-0.97,-0.13,-0.20)$ obtained for $w_{acc} = 0.5$, $w_{acc} = 0.75$, and $w_{acc} = 1.0$. Table \ref{tab_planWt50} summarizes the plan where different combinations of fixtures 2,3, and 4 and tools 1, 2, and 3 are utilized. As expected, increasing $w_{acc}$ results in a near-net shape with higher more support removability, i.e., lower discrepancy between the as-designed and as-manufactured model. For this build orientation, 90\% of the supports have been removed. 

 \begin{table} [h!]
	\caption{Summary of results for plan for the optimized build orientation  $\buildDir^*_1 = (-0.97,-0.13,-0.20)$ at $w_{acc} = 0.50$ with 90\% removable supports.}
	\tabulinesep=1mm
	\begin{tabu}[t!]{lcccl}
		\hline \hline 
		Step &Vol. Frac. (\%) & Fixture  & Tool& Direction\\
		\hline
		1 & 64.52 & 4 & 2 & (1,0,0)\\
		2 & 13.17& 4 & 1 & (-1,1,0)\\
		3 & 6.14 & 3 & 3 & (-1,0,-1)\\
		4 & 1.80 & 4 & 2 & (0,-1,-1)\\
		5 & 1.67 & 2 & 3 & (1,1,0)\\
		6 & 1.14 & 4 & 3& (1,0,1)\\
		7 & 0.92 & 3 & 2& (0,1,-1)\\
		8 & 0.87 & 3 & 2& (-1,0,0)\\
		\hline
	\end{tabu}
	\label{tab_planWt50}
\end{table}

\section{Conclusion}

We presented a framework to find the optimal build direction for a special case of hybrid manufacturing -- additive manufacturing followed by machining processes, where the additive process fabricates a near-net shape comprising the final part and the sacrificial support structures. Assuming the supports need to be removed using multi-axis machining, we need to ensure that the build orientation is chosen such that the support are ``accessible''. Our methodology is based on the IMF, which incorporates well-established concepts from  spatial planning and robotics to define a continuous field that measures minimum collision possible between an obstacle and cutting tools to come in contact with a given query point in the Euclidean space. We have extended IMF to near-net shapes (part and support) to find accessible and secluded regions of support for a given build orientation. Further, we have generalized IMF definition to multiple fixturing devices and configurations which enables more accurate analysis for real-world industrial applications.

In this paper, we focused on finding the best build orientation w.r.t. total support volume and secluded support volume. Since the total support volume depends on the build orientation and the secluded support volume depends on the given multi-axis machining setup, lower support volume does not necessarily result in lower secluded volume. In other words, the two objectives are often competing, thus the proposed algorithm finds the Pareto-optimal orientation to minimize both volumes, the former reducing the cost of materials and AM build time while the latter reduces the cost and time of SM post-processing. The global optimum will depend on the cost model for AM and SM (not considered in this paper) and the Pareto frontier  provides the engineers with the flexibility to select it based on their use-cases.

We have also proposed an automated support removal planning algorithm based on IMF, where a sequence of actions for iteratively machining support volumes is provided. The planning algorithm is based on the maximal removable volume, where at each step we find a combination of fixturing device, tool and orientation that remove as much material as possible in a greedy fashion.   

For ease of implementation we chose to  sample the 3D orientation space and selected the Pareto-optimal solution without any further optimization. However, it is possible to improve the current framework through various optimization methods, such as gradient-free or stochastic methods. In practical applications, the optimal plan must also take contact forces into account and considering maximal removable volume may not yield the best sequence of actions. 

The efficacy of the proposed method has been demonstrated through 2D and 3D examples. However, it has also been shown that considering the support removal solely at the post-processing stage is inherently limited and can pose many challenges where there might not exist a build orientation at which all supports are removable. In such cases, a time consuming and costly trial-and-error cycle must be repeated to modify the design. Future work will focus on incorporating support accessibility constraint upfront in the early stage of design and coupling design optimization with the build orientation optimization proposed in this paper.


\bibliographystyle{elsarticle-num} 
\bibliography{mfgPlanSuppRem}

\end{document}